\documentclass[a4paper]{jpconf}
\usepackage{amsmath}
\usepackage{graphicx}
\pdfoutput=1

\begin{document}

\title{Does the Heisenberg uncertainty principle apply along the time dimension?}
\author{John Ashmead}
\address{Visiting Scholar, University of Pennsylvania}

\ead{jashmead@seas.upenn.edu}

\begin{abstract}
Does the Heisenberg uncertainty principle (HUP) apply along the time
dimension in the same way it applies along the three space dimensions?
Relativity says it should; current practice says no. With recent advances
in measurement at the attosecond scale it is now possible to decide
this question experimentally. 

The most direct test is to measure the time-of-arrival of a quantum
particle: if the HUP applies in time, then the dispersion in the time-of-arrival
will be measurably increased. 

We develop an appropriate metric of time-of-arrival in the standard
case; extend this to include the case where there is uncertainty in
time; then compare. There is -- as expected -- increased uncertainty
in the time-of-arrival if the HUP applies along the time axis. The
results are fully constrained by Lorentz covariance, therefore uniquely
defined, therefore falsifiable.

So we have an experimental question on our hands. Any definite resolution
would have significant implications with respect to the role of time
in quantum mechanics and relativity. A positive result would also
have significant practical applications in the areas of quantum communication,
attosecond physics (e.g. protein folding), and quantum computing.
\end{abstract}

\section{Introduction}

\label{sec:introduction}
\begin{quotation}
``You can have as much junk in the guess as you like, provided that
the consequences can be compared with experiment.'' -- Richard P.
Feynman \cite{Feynman:1965jb}
\end{quotation}

\paragraph{Heisenberg uncertainty principle in time}

The Heisenberg uncertainty principle in space:

\begin{equation}
\Delta x\Delta p\geq1
\end{equation}

is a foundational principle of quantum mechanics. From special relativity
and the requirement of Lorentz covariance we expect that this should
be extended to time:

\begin{equation}
\Delta t\Delta E\geq1
\end{equation}

Both Bohr and Einstein regarded the uncertainty principle in time
as essential to the integrity of quantum mechanics.

At the sixth Conference of Solvay in 1930, Einstein devised the celebrated
Clock-in-a-Box experiment to refute quantum mechanics by showing the
HUP in time could be broken. Consider a box full of photons with a
small fast door controlled by a clock. Weigh the box. Open the door
for a time $\Delta t$, long enough for one photon to escape. Weigh
the box again. From the equivalence of mass and energy, the change
in the weight of the box gives you the \emph{exact} energy of the
photon. If $\Delta E\to0$, then you can make $\Delta t\Delta E<1$. 

Bohr famously refuted the refutation by looking at the specifics of
the experiment. If you weigh the box by adding and removing small
weights till the box's weight is again balanced, the process will
take time. During this process the clock will move up and down in
the gravitational field, and its time rate will be affected by the
gravitational redshift. Bohr showed that resulting uncertainty is
precisely what is required to restore the HUP in time \cite{Schilpp:1949oz,Pais:1982aa}.
The irony of employing Einstein's own General Relativity (via the
use of the gravitational redshift) to refute Einstein's argument was
presumably lost on neither Einstein nor Bohr.

In later work this symmetry between time and space was in general
lost. This is clear in the Schrödinger equation:

\begin{equation}
\imath\frac{\partial}{\partial\tau}\psi_{\tau}\left(\vec{x}\right)=H\psi_{\tau}\left(\vec{x}\right)\label{eq:schroedinger-equation}
\end{equation}

Here the wave function is indexed by time: if we know the wave function
at time $\tau$ we can use this equation to compute the wave function
at time $\tau+\epsilon$. The wave function has in general non-zero
dispersion in space, but is always treated as having zero dispersion
in time. In quantum mechanics ``time is a parameter not an operator''
(Hilgevoord \cite{Hilgevoord:1996bh,Hilgevoord:1998qu}). Or as Busch
\cite{Busch-2001} puts it ``\ldots{} different types of time energy
uncertainty can indeed be deduced in specific contexts, but \ldots{}
there is no unique universal relation that could stand on equal footing
with the position-momentum uncertainty relation.'' See also Pauli,
Dirac, and Muga \cite{Pauli:1980wd,Dirac:1958ty,Muga:2002ft,Muga:2008vv}.

So, we have a contradiction at the heart of quantum mechanics. Symmetry
requires that the uncertainty principle apply in time; experience
to date says this is not necessary. Both are strong arguments; neither
is decisive on its own. The goal of this work is to reduce this to
an experimental question; to show we can address the question with
current technology.

\paragraph{Order of magnitude estimate }

The relevant time scale is the time for a photon to cross an atom.
This is of order attoseconds. This is small enough to explain why
such an effect has not already been seen by chance. Therefore the
argument from experience is not decisive.

But recent advances in ultra-fast time measurements have been extraordinary;
we can now do measurements at even the sub-attosecond time scale \cite{Ossiander:2016fp}.
Therefore we should now be able to measure the effects of uncertainty
in time, if they are present.

Further the principle of covariance strongly constrains the effects.
In the same way that the inhabitants of Abbott's Flatland \cite{Abbott:1884fn}
can infer the properties of a Sphere from the Sphere's projection
on Flatland, we should be able to predict the effects of uncertainty
in time from the known uncertainties in space. 

\paragraph{Operational meaning of uncertainty in time}

In an earlier work \cite{Ashmead_2019} we used the path integral
approach to do this\footnote{Detailed references are provided in the earlier work. For here we
note that the initial ideas come from the work of Stueckelberg and
Feynman \cite{Stueckelberg:1941aa,Stueckelberg:1941la,Feynman:1948,Feynman:1949sp,Feynman:1949uy,Feynman:1950rj}
as further developed by Horwitz, Fanchi, Piron, Land, Collins, and
others \cite{Horwitz:1973ys,Fanchi:1978aa,Fanchi:1993aa,Fanchi:1993ab,Land:1996aj,Horwitz:2005ix,Fanchi:2011aa,Horwitz:2015jk}.}. We generalized the usual paths in the three space dimensions to
extend in time as well. No other change was required. The results
were manifestly covariant by construction, uniquely defined, and consistent
with existing results in the appropriate long time limit.

While there are presumably other routes to the same end, the requirement
of Lorentz covariance is a strong one and implies -- at a minimum
-- that the first order corrections of any such route will be the
same. We can therefore present a well-defined and falsifiable target
to the experimentalists.

\paragraph{Time-of-arrival measurements}

The most obvious line of attack is to use a time-of-arrival detector.
We emit a particle at a known time, with known average velocity and
dispersion in velocity. We measure the dispersion in time at a detector.
The necessary uncertainty in space will be associated with uncertainty
in time-of-arrival. For instance if the wave function has an uncertainty
in position of $\Delta x$ and an average velocity of $v$, there
will be an uncertainty of time-of-arrival of order $\frac{\Delta x}{v}$.
We refer to this as the extrinsic uncertainty.

If there is an intrinsic uncertainty in time associated with the particle
we expect to see an additional uncertainty in time-of-arrival from
that. If the intrinsic uncertainty in time is $\Delta t$, then we
would expect a total uncertainty in time-of-arrival of order $\Delta t+\frac{\Delta x}{v}$.
It is the difference between the two predictions that is the experimental
target.

For brevity we will refer to standard quantum mechanics, without uncertainty
in time, as SQM. And we will refer to quantum mechanics with intrinsic
uncertainty in time as TQM.

\paragraph{Time-of-arrival in SQM}

We clearly need a solid measure of the extrinsic uncertainty to serve
as a reference point. We look at a number of available measures but
are forced to conclude none of the established metrics are entirely
satisfactory: they impose arbitrary conditions only valid in classical
mechanics, have free parameters, or are physically unrealistic. We
argue that the difficulties are of a fundamental nature. In quantum
mechanics there cannot really be a crisp boundary between the wave
function detected and the equipment doing the detection: we must --
as a Bohr would surely insist -- consider both simultaneously as
part of a single system. However once this requirement is recognized
and accepted, we can work out the rules for a specific setup and get
reasonably solid predictions.

\paragraph{Comparison of time-of-arrival in SQM and TQM}

With this preliminary question dealt with, the actual comparison of
the results with and without intrinsic uncertainty in time is straightforward.
We first look at a generic particle detector setup in SQM, then look
at the parallel results in TQM. By writing the latter as the direct
product of a time part and a space part, we get a clean comparison.
As expected, if there is intrinsic uncertainty in time then the uncertainty
in time-of-arrival is increased and by a specific amount.

In the non-relativistic limit, the relative increase in the uncertainty
in time-of-arrival is small. But by running the initial wave function
through a single slit in time, as a rapidly opening and closing camera
shutter, we can make the relative increase arbitrarily great.

\emph{We therefore have falsifiability.}

\paragraph{Implications}

Since the predictions come directly from the principle of Lorentz
covariance and the basic rules for quantum mechanics, any definite
result -- negative or positive -- will have implications for foundational
questions in relativity and quantum mechanics.

If there is in fact intrinsic uncertainty in time, there will be practical
implications for quantum communication, attosecond physics (e.g. protein
folding), and quantum computing. This will open up new ways to carry
information, additional kinds of interaction between quantum particles
(forces of anticipation and regret, resulting from the extension in
time), and a deeper understanding of the measurement problem.

\section{Time-of-arrival measurements in SQM}

\label{sec:SQM}

There does not appear to be a well-defined, generally accepted, and
appropriate metric for the time-of-arrival in the case of SQM.

The Kijowski metric \cite{Kijowski:1974hi} is perhaps the most popular
metric for time-of-arrival. It has a long history, back to 1974, has
seen a great deal of use, and has some experimental confirmation.
However it is based on a deeply embedded classical assumption which
is unjustifiable from a quantum mechanical viewpoint and which produces
inconsistent results in some edge cases. We will nevertheless find
it useful as a starting point.

A conceptually sounder approach is supplied by Marchewka and Schuss,
who use Feynman path integrals to compute the time-of-arrival \cite{Marchewka:1998aa,Marchewka:1999db,Marchewka:1999nl,Marchewka:2000ys}.
Unfortunately their approach has free parameters, which rules it out
for use here.

The difficulties with the Marchewka and Schuss approach are partly
a function of an ad hoc treatment of the boundary between incoming
wave function and the plane of the detector. We are able to deal with
this boundary in a systematic way by using a discrete approach along
the space coordinate, then taking the continuum limit.

Unfortunately while this discrete approach solves the problem of how
to handle the discontinuity at the edge of the detector it does not
address the more fundamental problem: of assuming a crisp boundary
between the quantum mechanical and classical parts of the problem
in the first place.

We address this problem by arguing that while the assumption of a
crisp boundary between the quantum mechanical and classical parts
of a problem is not in general justified it is also not in general
necessary. If we work bottom-up, starting from a quantum mechanical
perspective, we can use various ad hoc but reasonable heuristics to
determine which parts may be described with acceptable accuracy using
classical mechanics, and where only a fully quantum mechanical approach
will suffice.

\subsection{Kijowski time-of-arrival operator}
\begin{quote}
``A common theme is that classical mechanics, deterministic or stochastic,
is always a fundamental reference on which all of these approaches
{[}to the time-of-arrival operator{]} are based.'' -- Muga and Leavens
\cite{Muga:2000nx}
\end{quote}

\subsubsection{Definition}

The most popular time-of-arrival operator seems to be one developed
by Kijowski in 1974 \cite{Kijowski:1974hi}. It and variations on
it have seen much use since then \cite{Baute:2000ab,Baute:2000aa,Muga:2000nx,Baute:2001aa,Ruggenthaler:2005aa,Anastopoulos:2006ab,Yearsley:2010ac,Kiukas:2011aa,Yearsley:2011ab,Halliwell:2015aa,Kijowski:2015aa,Das:2018fk,Yearsley:2011ac}.
We used Kijowski as a starting point in our previous paper: it gives
reasonable results in simple cases. Unfortunately its conceptual problems
rule it out for use here. 

The main problem is that classical requirements play an essential
role in Kijowski's original derivation. In particular, Kijowski assumed
that for the part of the wave function to the left of the detector\footnote{Throughout we assume that the particle is going left to right along
the $x$ axis; that the detector is at $x=0$; that the particle starts
well to the left of the detector; and that we do not need to consider
the $y$ and $z$ axes. These assumptions are conventional in the
time-of-arrival literature.} only components with $p>0$ will contribute; while for the part of
the wave function to the right of the detector only components with
$p<0$ will contribute. This condition (which we will refer to as
the classical condition\label{classical-condition}) is imposed on
the basis of the corresponding requirement in classical mechanics,
as noted by Muga and Leavens above.
\begin{figure}
\includegraphics[scale=0.75]{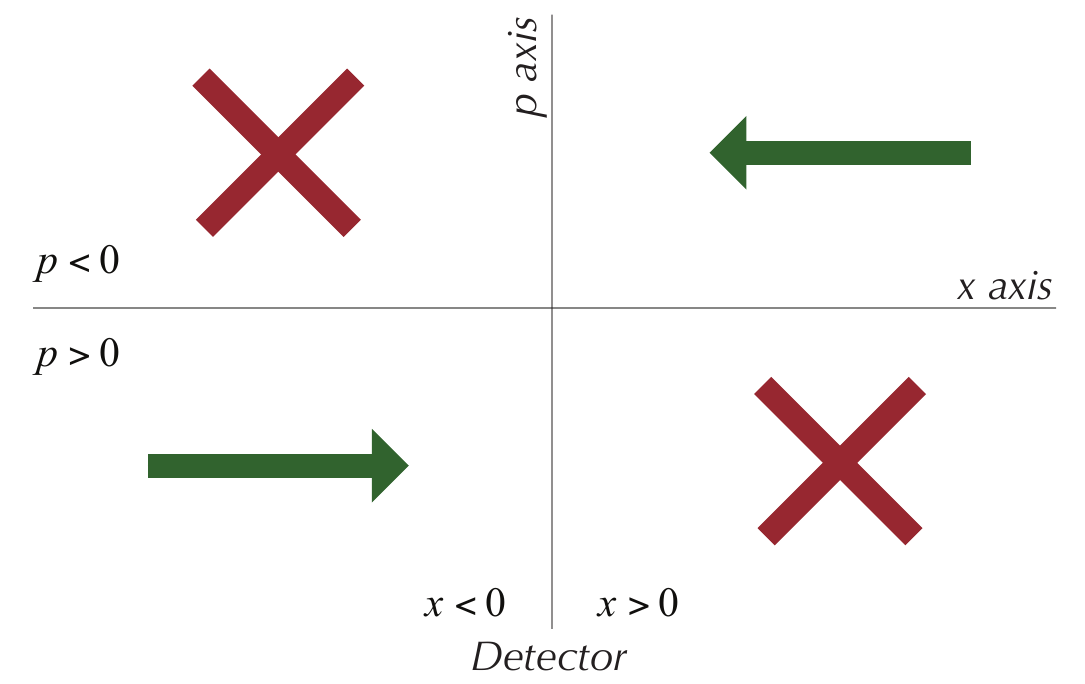}

\caption{The classical condition for time-of-arrival measurements}
\end{figure}

Focusing on the case of a single space dimension, with the detector
at $x=0$, the Kijowski metric is\footnote{Equation 122 in \cite{Muga:2000nx}. }:

\begin{equation}
\rho_{d}\left(\tau\right)={\left|{\int\limits _{0}^{\infty}dp\sqrt{\frac{p}{{2\pi m}}}{e^{-\imath\frac{{{p^{2}}\tau}}{{2m}}}}{\varphi^{\left({left}\right)}}\left(p\right)}\right|^{2}}+{\left|{\int\limits _{-\infty}^{0}dp\sqrt{\frac{{-p}}{{2\pi m}}}{e^{-\imath\frac{{{p^{2}}\tau}}{{2m}}}}{\varphi^{\left({right}\right)}}\left(p\right)}\right|^{2}}
\end{equation}

However while the restrictions on momentum may make sense classically\footnote{Amusingly enough, this condition is sometimes violated even in classical
waves: Rayleigh surface waves in shallow water (e.g. tsunamis approaching
land) show retrograde motion of parts of the wave \cite{Rayleigh:1885aa}.}, they are more troubling in quantum mechanics. It is as if we were
thinking of the wave function as composed of a myriad of classical
particles each of which travels like a tiny billiard ball. This would
appear to be a hidden variable interpretation of quantum mechanics.
This was shown to be inconsistent with quantum mechanics by Bell in
1965 \cite{Bell:1964aa} experimentally confirmed \cite{Clauser:1969aa,Aspect:1981aa,Aspect:1982aa,Aspect:1982ab,Hensen:2015aa}.

However, Kijowski's metric has produced reasonable results, see for
instance \cite{Muga:2000aa}. Therefore if we see problems with it,
we need to explain the successes as well.

\subsubsection{Application to Gaussian test functions}

We will use Gaussian test functions (GTFs) to probe this metric. We
define GTFs as normalized Gaussian functions that are solutions to
the free Schrödinger equation. We will look at the results of applying
Kijowski's metric to two different cases, which we refer to as the
``bullet'' and ``wave'' variations, following Feynman's terminology
in his discussion of the double-slit experiment \cite{Feynman:1965ah}.
By ``bullet'' we mean a wave function which is narrowly focused
around the central axis of motion; by ``wave'' we mean a wave function
which is widely spread around the central axis of motion.

\paragraph{Bullets}

For the bullet version we will take a GTF which starts at time $\tau=0$,
centered at start position $x_{0}=-d$, with average momentum $p_{0}$,
and dispersion $\sigma_{p}^{2}\ll p_{0}$. These conditions are met
in the Muga, Baute, Damborenea, and Egusquiza \cite{Muga:2000aa}
reference; they give:

\begin{equation}
\delta v=0.098{{cm}\mathord{\left/{\vphantom{{cm}s}}\right.\kern-\nulldelimiterspace}s},{v_{0}}=10{{cm}\mathord{\left/{\vphantom{{cm}s}}\right.\kern-\nulldelimiterspace}s}\label{eq:muga-test-case}
\end{equation}

This is typical of high-energy particles, where the free particle
wave functions will have essentially no negative momentum component.
The starting wave function is:

\begin{equation}
{{\bar{\varphi}}_{0}}\left(x\right)=\sqrt[4]{{\frac{1}{{\pi\sigma_{x}^{2}}}}}{e^{\imath{p_{0}}x-\frac{1}{{2\sigma_{x}^{2}}}{{\left({x+d}\right)}^{2}}}}\label{eq:bullet-wave-function}
\end{equation}

This is taken to satisfy the free Schrödinger equation:

\begin{equation}
\imath\frac{\partial}{{\partial\tau}}{{\bar{\varphi}}_{\tau}}\left(x\right)=-\frac{1}{{2m}}\frac{{\partial^{2}}}{{\partial{x^{2}}}}{{\bar{\varphi}}_{\tau}}\left(x\right)\label{eq:schrodinger-equation-1}
\end{equation}

so is given as a function of time by:

\begin{equation}
\bar{\varphi}_{\tau}\left(x\right)=\sqrt[4]{\frac{1}{\pi\sigma_{x}^{2}}}\sqrt{\frac{1}{f_{\tau}^{\left(x\right)}}}e^{\imath p_{0}x-\frac{1}{2\sigma_{x}^{2}f_{\tau}^{\left(x\right)}}\left(x+d-\frac{p_{0}}{m}\tau\right)^{2}-\imath\frac{p_{0}^{2}}{2m}\tau}\label{eq:gtf-in-space}
\end{equation}

with dispersion factor $f$ defined by:

\begin{equation}
f_{\tau}^{\left(x\right)}\equiv1+\imath\frac{\tau}{m\sigma_{x}^{2}}\label{eq:dispersion-factor}
\end{equation}

The momentum transform is:

\begin{equation}
{{\hat{\bar{\varphi}}}_{\tau}}\left(p\right)=\sqrt[4]{{\frac{1}{{\pi\sigma_{p}^{2}}}}}{e^{\imath pd-\frac{{{\left({p-{p_{0}}}\right)}^{2}}}{{2\sigma_{p}^{2}}}-\imath\frac{{p^{2}}}{{2m}}\tau}}\label{eq:momentum-distribution}
\end{equation}

with $\sigma_{p}\equiv\frac{1}{\sigma_{x}}$. We will use this wave
function as a starting point throughout this work, referring to it
as the ``bullet wave function''.

In Kijowski's notation, there is no right hand wave function since
the particle started on the left. Further since the wave function
is by construction narrow in momentum space we can extend the lower
limit in the integral on the left from zero to negative infinity giving:

\begin{equation}
{\rho_{d}}\left(\tau\right)={\left|{\int\limits _{-\infty}^{\infty}dp\sqrt{\frac{p}{{2\pi m}}}\sqrt[4]{{\frac{1}{{\pi\sigma_{p}^{2}}}}}{e^{\imath pd-\frac{{{\left({p-{p_{0}}}\right)}^{2}}}{{2\sigma_{p}^{2}}}-\imath\frac{{p^{2}}}{{2m}}\tau}}}\right|^{2}}
\end{equation}

Since the wave function is narrowly focused, we can replace the square
root in the integral by the average: 

\begin{equation}
\sqrt{\frac{{{p_{0}}+\delta p}}{m}}\approx\sqrt{\frac{{{\text{p}}_{0}}}{m}}
\end{equation}

which we can then pull outside of the integral:

\begin{equation}
{\rho_{d}}\left(\tau\right)\approx\frac{{p_{0}}}{m}{\left|{\int\limits _{-\infty}^{\infty}{\frac{{dp}}{{\sqrt{2\pi}}}\sqrt[4]{{\frac{1}{{\pi\sigma_{p}^{2}}}}}{e^{\imath pd-\frac{{{\left({p-{p_{0}}}\right)}^{2}}}{{2\sigma_{p}^{2}}}-\imath\frac{{p^{2}}}{{2m}}\tau}}}}\right|^{2}}
\end{equation}

We recognize the integral as simply the inverse Fourier transform
of the momentum space form: 

\begin{equation}
{{\bar{\varphi}}_{\tau}}\left(d\right)=\sqrt[4]{{\frac{1}{{\pi\sigma_{x}^{2}}}}}\sqrt{\frac{1}{{f_{\tau}^{\left(x\right)}}}}{e^{\imath{p_{0}}d-\frac{1}{{2\sigma_{x}^{2}f_{\tau}^{\left(x\right)}}}{{\left({d-\frac{{p_{0}}}{m}\tau}\right)}^{2}}-\imath\frac{{p_{0}^{2}}}{{2m}}\tau}}
\end{equation}

so:

\begin{equation}
{\rho_{d}}\left(\tau\right)\approx\frac{{p_{0}}}{m}{\left|{{{\bar{\varphi}}_{\tau}}\left(d\right)}\right|^{2}}=\frac{{p_{0}}}{m}{{\bar{\rho}}_{\tau}}\left(d\right)\label{eq:toa-sqm-kijowski-blob}
\end{equation}

with the probability density being:

\begin{equation}
{{\bar{\rho}}_{\tau}}\left(d\right)=\frac{{p_{0}}}{m}\sqrt{\frac{1}{{\pi\sigma_{x}^{2}\left({1+\frac{{\tau^{2}}}{{{m^{2}}\sigma_{x}^{4}}}}\right)}}}\exp\left({-\frac{{{\left({{\bar{x}}_{\tau}}\right)}^{2}}}{{\sigma_{x}^{2}\left({1+\frac{{\tau^{2}}}{{{m^{2}}\sigma_{x}^{4}}}}\right)}}}\right)\label{eq:detection-rate}
\end{equation}

and with the average location in space:

\begin{equation}
\bar{x}_{\tau}\equiv-d+{v_{0}}\tau
\end{equation}

We have the probability density as a function of $x$; we need it
as a function of $\tau$. We therefore make the variable transformation:

\begin{equation}
d\tau={v_{0}}dx=\frac{{p_{0}}}{m}dx\label{eq:toa-sqm-kijowski-tau-space}
\end{equation}

And rewrite $\tau$ as the average value of $\tau$ plus an offset:

\begin{equation}
\tau=\bar{\tau}+\delta\tau,\bar{\tau}\equiv\frac{d}{{v_{0}}}
\end{equation}

We rewrite the numerator in the density function in terms of $\delta\tau$:

\begin{equation}
\bar{\rho}_{\tau}\left(d\right)\approx\sqrt{\frac{1}{\pi\sigma_{x}^{2}\left|f_{\tau}^{\left(x\right)}\right|^{2}}}e^{-\frac{\left(v_{0}\delta\tau\right)^{2}}{\sigma_{x}^{2}\left|f_{\tau}^{\left(x\right)}\right|^{2}}}\label{eq:kijowski-dispersion-in-time}
\end{equation}

We may estimate the uncertainty in time by expanding around the average
time. Since the numerator is already only of second order in $\delta\tau$
we need only keep the zeroth order in $\delta\tau$ in the denominator:
\begin{equation}
\sigma_{x}^{2}\left|f_{\tau}^{\left(x\right)}\right|^{2}=\sigma_{x}^{2}+\frac{\tau^{2}}{m^{2}\sigma_{x}^{2}}\approx\frac{\tau^{2}}{m^{2}\sigma_{x}^{2}}=\frac{\left(\bar{\tau}+\delta\tau\right)^{2}}{m^{2}\sigma_{x}^{2}}\approx\frac{\bar{\tau}^{2}}{m^{2}\sigma_{x}^{2}}\label{eq:probdens-space}
\end{equation}
giving: 
\begin{equation}
\bar{\rho}_{\delta\tau}\approx\sqrt{\frac{v_{0}^{2}m^{2}\sigma_{x}^{2}}{\pi\bar{\tau}^{2}}}e^{-\frac{v_{0}^{2}m^{2}\sigma_{x}^{2}}{\bar{\tau}^{2}}\left(\delta\tau\right)^{2}}
\end{equation}
We define an effective dispersion in time: 
\begin{equation}
\bar{\sigma}_{\tau}\equiv\frac{1}{mv_{0}\sigma_{x}}\bar{\tau}\label{eq:bullet-dispersion}
\end{equation}

So we have the probability of detection as roughly: 
\begin{equation}
\bar{\rho}_{\delta\tau}\approx\sqrt{\frac{1}{\pi\bar{\sigma}_{\tau}^{2}}}e^{-\frac{\left(\delta\tau\right)^{2}}{\bar{\sigma}_{\tau}^{2}}}\label{eq:probdens-space-1}
\end{equation}

This is normalized to one, centered on $\tau=\bar{\tau}$, and with
uncertainty:
\begin{equation}
\Delta\tau=\frac{1}{\sqrt{2}}\bar{\sigma}_{\tau}\label{eq:kijowski-bullet-uncertainty}
\end{equation}

It is interesting that the classical condition has played no essential
role in this derivation. In spite of that, the result is about what
we would expect from a classical particle with probability distribution
$\bar{\rho}$. Therefore the implication of the experimental results
in Muga and Leavens is that they are confirmation of the reasonableness
of the classical approximation in general -- rather than of Kijowski's
classical condition in particular.

\paragraph{Waves}

Now consider a more wave-like Gaussian test function: slow and wide.
Take $0<{p_{0}}\ll{\sigma_{p}}$ in equation \ref{eq:momentum-distribution},
in fact make $p_{0}$ only very slightly greater than 0. 

Our GTF is still moving to the right but very slowly. If we wait long
enough -- and if there were no detector in the way -- it will eventually
get completely past the detector. Since there is a detector, the GTF
will encounter this instead. We will first assume a perfect detector
with no reflection.

Use of the Kijowski distribution requires we drop the $p<0$ half
of the wave function because we started on the left:

\begin{equation}
{\varphi^{\left({wave}\right)}}\left(p\right)\Rightarrow\sqrt[4]{{\frac{1}{{\pi\sigma_{p}^{2}}}}}{e^{\imath pd-\frac{{\left({p-{p_{0}}}\right)^{2}}}{{2\sigma_{p}^{2}}}}}\theta\left(p\right)
\end{equation}

We can no longer justify extending the lower bound on the left hand
integral from zero to negative infinity. We have instead:

\begin{equation}
{\rho_{d}}\left(\tau\right)={\left|{\int\limits _{0}^{\infty}dp\sqrt{\frac{p}{{2\pi m}}}\sqrt[4]{{\frac{1}{{\pi\sigma_{p}^{2}}}}}{e^{\imath pd-\frac{{{\left({p-{p_{0}}}\right)}^{2}}}{{2\sigma_{p}^{2}}}-\imath\frac{{p^{2}}}{{2m}}\tau}}}\right|^{2}}
\end{equation}

This means we have dropped nearly half of the wave function and therefore
expect a detection rate of only about 25\%. 

The integral over the momentum gives a sum over Bessel functions of
fractional order; the subsequent integrals over time are distinctly
non-trivial. However in the limiting case $d\to0,p_{0}\to0$ (the
wave function starts at the detector) the distribution simplifies
drastically giving:

\begin{equation}
\rho_{0}\left(\tau\right)={\left|{\frac{{\sqrt[4]{m}\sqrt{\sigma_{p}^{2}}\Gamma\left({\frac{3}{4}}\right)}}{{{{(2\pi)}^{3/4}}{{(m+\imath\sigma_{p}^{2}\tau)}^{3/4}}}}}\right|^{2}}
\end{equation}

And this in turn is simple enough to let us compute the norm analytically: 

\begin{equation}
\int\limits _{0}^{\infty}{d\tau\rho_{0}\left(\tau\right)}=\frac{1}{4}
\end{equation}

This is exactly as expected: if we throw out half the wave function,
the subsequent norms are the square of the wave function or one quarter.

This result is general. The GTF obeys the Schrödinger equation, which
is norm preserving. If we throw out fraction $F$ of the wave function,
the resulting wave function has norm ${\left({1-F}\right)^{2}}$.
Assuming perfect detection, the Kijowski distribution will under-count
the actual detections by this factor.

It is arguable that the problem is the assumption of perfect detection.
However if we drill down to the actual mechanism of detection, this
will involve some sort of interaction of the incoming particles with
various atoms, typically involving the calculation of matrix elements
of the form (e.g. Sakurai \cite{Sakurai:1967tl}):

\begin{equation}
\left\langle {\varphi_{\tau}^{\left({n'l'm'}\right)}\left({\vec{x}}\right)}\right|\sqrt{\frac{{n_{\vec{k},\alpha}}}{{2\omega V}}}{{\vec{\varepsilon}}^{\left(\alpha\right)}}\exp\left({\imath\vec{k}\cdot\vec{x}-\imath\omega\tau}\right)\left|{\varphi_{\tau}^{\left({nlm}\right)}\left({\vec{x}}\right)}\right\rangle 
\end{equation}

It is difficult to make sense of the classical condition in this context.
Would we have to apply it to the interaction with each atom in turn?
And how would the atoms know whether they are supposed to be detecting
particles coming from the left or the right? And therefore how would
the unfortunate atoms know whether they should drop the upper left
and bottom right quadrants on the diagram or vice versa?

\subsubsection{Requirements for a time-of-arrival metric}

\label{subsec:Requirements}

The classical condition was introduced without justification in terms
of quantum mechanics itself. As there are no cases where the rules
of quantum mechanics have failed to produce correct results, this
is not acceptable.

Further it is unnecessary. The probability current, discussed below
(subsubsection \ref{subsec:Probability-current-sqm}), provides a
conceptually sound and purely quantum mechanical approach to the problem
of detection. 

We therefore have to dispense entirely with the classical condition.
As this is the foundation of the Kijowski class of metrics -- per
Muga and Leavens above -- we have to dispense with the class as well.

We have however learned a bit about our requirements: we are looking
for a time-of-arrival metric which satisfies the requirements:
\begin{enumerate}
\item Comes out of a completely quantum mechanical analysis, with no ad
hoc classical requirements imposed on it.
\item Includes an account of the interaction with the detector. As Bohr
pointed out during the EPR debate \cite{Einstein:1935er,Bohr:1935bt},
an analysis of a quantum mechanical detection must include the specifics
of the apparatus. There is no well-defined value of a quantum parameter
in the abstract, but only in the context of a specific measurement.
\item Conserves probability.
\item And of course, is as simple as possible given the first three requirements.
\end{enumerate}

\subsection{Marchewka and Schuss path integral approach}

\label{subsec:Marchewka-and-Schuss}
\begin{quote}
``According to our assumptions, trajectories that propagate into
the absorbing boundary never return into the interval {[}a, b{]} so
that the Feynman integral over these trajectories is supported outside
the interval. On the other hand, the Feynman integral over the bounded
trajectories in the interval is supported inside the interval. Thus
the two integrals are orthogonal and give rise to no interference.''
-- Marchewka and Schuss \cite{Marchewka:1999nl}
\end{quote}
We turn to an approach from Marchewka and Schuss \cite{Marchewka:1998aa,Marchewka:1999db,Marchewka:1999nl,Marchewka:2000ys}.
They use a path integral approach plus the assumption of an absorbing
boundary. To compute the wave function, they use a recursive approach:
computing the wave function at each step, subtracting off the absorption
at each step, and so to the end. 

Marchewka and Schuss start with the wave function defined over a general
interval from $a\to b$. We can simplify their treatment slightly
by taking $a=-\infty,b=0$. They break the time evolution up into
steps of size $\epsilon$. Given the wave function at step $n$, they
compute the wave function at step $n+1$ as: 

\begin{equation}
{\psi_{n+1}}\left({x_{n+1}}\right)=\int\limits _{-\infty}^{0}{d{x_{n}}{K_{\varepsilon}}\left({{x_{n+1}};{x_{n}}}\right)}{\psi_{n}}\left({x_{n}}\right)
\end{equation}

The kernel can be quite general.

Marchewka and Schuss then introduce the assumption that the probability
to be absorbed at each endpoint is proportional to the value of the
probability to be at the end point times a characteristic length $\lambda$\footnote{They refer to $\lambda$ as a ``fudge parameter''. The value of
$\lambda$ has to found by running some reference experiments; after
this is done the values are determined for ``nearby'' experiments.}:

\begin{equation}
{P_{n}}=\lambda{\left|{{\psi_{n}}\left(0\right)}\right|^{2}}
\end{equation}

Using the kernel they compute the probability of absorption as:

\begin{equation}
{P_{n}}=\frac{\varepsilon}{{2\pi{m}}}\left({\lambda\left|{\left.{{\partial_{x}}{\psi_{n}}\left(x\right)}\right|_{x=0}^{2}}\right|}\right)
\end{equation}

They then shrink the wave function at the next step by an appropriate
factor, the survival probability:

\begin{equation}
{\psi_{n+1}}\left(x\right)\to\sqrt{1-{P_{n}}}{\psi_{n+1}}\left(x\right)
\end{equation}

This guarantees the total probability is constant, where the total
probability is the probability to be detected plus the probability
to survive. This overcomes the most obvious problem with the Kijowski
approach. 

At the same time it represents an explicit loss of the phase information.
There is nothing to keep the ``characteristic length'' from being
complex; taking it as real is a choice. Therefore we are seeing the
loss of information normally associated with the act of measurement.
So we have effectively broken the total measurements at the end points
down into a series of time dependent ``mini-measurements'' or ``mini-collapses''
at each time step.

Computing the product of all these infinitesimal absorptions they
get the final corrected wave function as:

\begin{equation}
\psi_{\tau}^{\left({corrected}\right)}\left(x\right)=\exp\left({-\frac{1}{{2\pi{m}}}\int\limits _{0}^{\tau}{d\tau'\left({{\lambda}\left|{{\left.{{\partial_{x}}{\psi_{\tau}}\left(0\right)}\right|}^{2}}\right|}\right)}}\right){\psi_{\tau}}\left(x\right)
\end{equation}

They go on to apply this approach to a variety of cases, as GTFs going
from left to right.

The Marchewka and Schuss approach is a significant advance over the
Kijowski metric: it works at purely quantum mechanical level, it includes
the detector in the analysis, and it guarantees by construction overall
conservation of probability.

But at the same time the introduction of the characteristic length
is a major problem for falsifiability: how do we know whether the
same length should be used for SQM and for TQM?

And the trick for estimating the loss of probability at each step
implicitly assumes no interference in time. This is not a problem
for SQM but a point we need to keep in mind when including TQM.

We therefor further refine our requirements for a metric: we need
to handle any discontinuity at the detector in a well-defined way.

\subsection{Discrete approach }

\label{sec:Feynman-checkboard}

\begin{figure}
\begin{centering}
\includegraphics[scale=0.8]{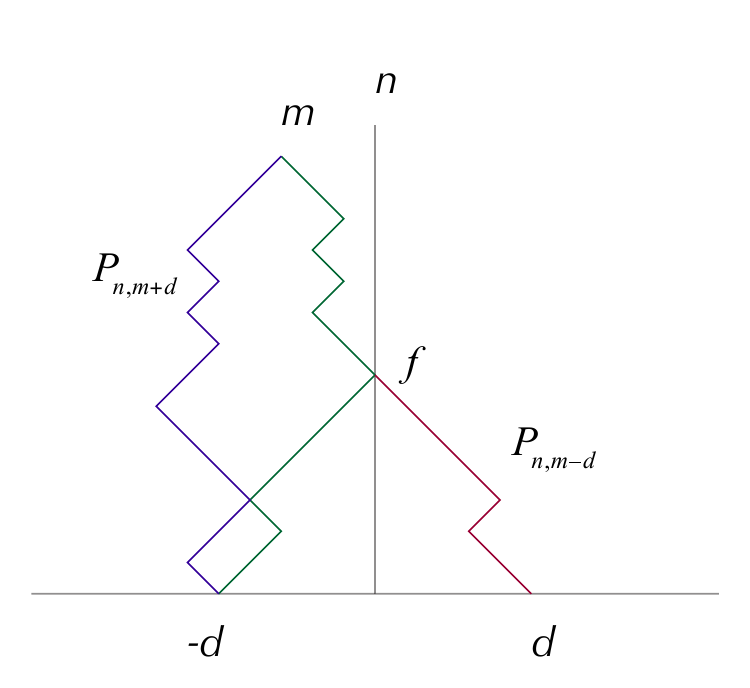}
\par\end{centering}
\caption{Reflection principle \label{fig:toa-rw-images}}
\end{figure}

We can handle the discontinuity at the boundary in a well-defined
way by taking a discrete approach, by setting up a grid in $\tau,x$.
We model quantum mechanics as a random walk on the Feynman checkerboard
\cite{Feynman:2010bt}. Feynman used steps taken at light speed; ours
will travel a bit more slowly than that. 

This lets us satisfy the previous requirements. And it lets us take
advantage of the extensive literature on random walks (e.g. \cite{Feller:1968he,Biggs:1989aa,Graham:1994ya,Ibe:2013aa}). 

We will start with a random walk model for detection, take the continuous
limit as the grid spacing goes to zero, Wick rotate in time, and then
apply the results to a GTF. This will give us a well-defined answer
for the detection rate for a GTF.

\subsubsection{Random Walk}

\label{subsec:Random-Walk}

We start with a grid, index by $n\in\left[{0,1,\ldots,\infty}\right]$
in the time direction and by $m\in[-\infty,\ldots,-1,0,1,\ldots,\infty]$
in the space direction. We start with a simple coin tossing picture.
We will take the probability to go left or right as equal, both $\frac{1}{2}$.
$P_{nm}$ is the probability to be at position $m$ at step $n.$
It is given recursively by:

\begin{equation}
{P_{n+1,m}}={P_{n,m-1}}/2+{P_{n,m+1}}/2\label{eq:rw-difference-equation}
\end{equation}

The number of paths from $\left(0,0\right)\to\left(n,m\right)$ is
given by the binomial coefficient $\left({\begin{array}{c}
n\\
{\frac{{n+m}}{2}}
\end{array}}\right)$. We will take this as zero if $n$ and $m$ do not have the same
parity, if they are not both even or both odd. If $n>\left|{m}\right|$
then $P_{nm}=0$. 

The total probability to get from $\left(0,0\right)\to\left(n,m\right)$
is given by the number of paths times the probability $\frac{{1}}{2^{n}}$
for each one. In our example:

\begin{equation}
{P_{nm}}=\left({\begin{array}{c}
n\\
{\frac{{n+m}}{2}}
\end{array}}\right)\frac{1}{{2^{n}}}\label{eq:rw-free-prob}
\end{equation}

If we wish to start at a position other than zero, say $m'$, we replace
$m$ with $m-m'$.

\paragraph{First arrival}

We can model detection as the first arrival of the path at the position
of the detector. We start the particle at step 0 at position $m=-d$
and position the detector at $m=0$. We take the probability of the
first arrival at step $n$ when starting at position $-d$ as ${F_{n}}$. 

We define the function $G_{nm}$ as the probability to arrive at position
$m$ at time $n$, \emph{without} having been previously detected.
We can get this directly from the raw probabilities using the reflection
principle (e.g. \cite{Feller:1968he}) as follows:

The number of paths that start at $\left(0,-d\right)$ and get to
$\left(n,m\right)$ is given by: $\left({\begin{array}{c}
n\\
{\frac{{n+{m+d}}}{2}}
\end{array}}\right)$. Now consider all paths that get to $m$ at step $n$ having first
touched the detector at some step $f$. By symmetry, the number of
paths from $-d\to f$ is the same as the number of paths from $d\to f$.
The number of these paths is $\left({\begin{array}{c}
n\\
{\frac{{n+\left|{m-d}\right|}}{2}}
\end{array}}\right)$. 

Therefore we have the number of paths that reach $m$ without first
touching the line $m=0$ as (\cite{Grimmett:2001aa}): 

\begin{equation}
{G_{nm}}=\left({\left({\begin{array}{c}
n\\
{\frac{{n+m+d}}{2}}
\end{array}}\right)-\left({\begin{array}{c}
n\\
{\frac{{n+m-d}}{2}}
\end{array}}\right)}\right)\frac{1}{{2^{n}}}
\end{equation}

And of course, if $\left(n,m\right)$ do not have the same parity,
$G_{nm}=0$.

To get to the detector at step $n$, the particle must have been one
step to the left at the previous step. It will then have a probability
of $\frac{{1}}{2}$ to get the detector. So we have: 

\begin{equation}
{F_{n}}=\frac{{1}}{2}{G_{n-1,-1}}
\end{equation}

giving:

\begin{equation}
F_{n}=\frac{d}{n}\left({\begin{array}{c}
n\\
{\frac{{n+d}}{2}}
\end{array}}\right)\frac{{1}}{2^{n}}=\frac{d}{n}{P_{nd}}
\end{equation}

If the particle is already at position $0$ at step $0$ we declare
that it has arrived at time zero so we have:
\begin{equation}
F_{0}={\delta_{d0}}\label{eq:rw-limit-at-zero}
\end{equation}

\subsubsection{Diffusion equation}

\label{subsec:Continuum-limit}

The passage to the continuum limit is well understood (\cite{Grimmett:2001aa,Ibe:2013aa}).
Define:

\begin{equation}
\begin{gathered}\tau=n\Delta\tau\hfill\\
x=m\Delta x\hfill
\end{gathered}
\end{equation}

We get the diffusion equation if we take the limit so that both $\Delta x$
and $\Delta\tau$ become infinitesimal while we keep fixed the ratio:

\begin{equation}
{D_{0}}=\mathop{\lim}\limits _{\Delta\tau,\Delta x\to0}\frac{{{\left({\Delta x}\right)}^{2}}}{{2\Delta\tau}}
\end{equation}

This gives the diffusion equation for the probability $P$:

\begin{equation}
\frac{{\partial P}}{{\partial\tau}}={D_{0}}\frac{{{\partial^{2}}P}}{{\partial{x^{2}}}}
\end{equation}

To further simplify we take $D_{0}=\frac{1}{2}$:

\begin{equation}
\frac{{\partial{P_{\tau}}}}{{\partial\tau}}=\frac{1}{2}\frac{{\partial^{2}}}{{\partial{x^{2}}}}{P_{\tau}}
\end{equation}

If we start with the probability distribution ${P_{0}}\left(x-x'\right)=\delta\left({x-x'}\right)$
the probability distribution as a function of time is then:

\begin{equation}
{P_{\tau}}\left({x;x'}\right)=\frac{1}{{\sqrt{2\pi\tau}}}\exp\left({-\frac{{{\left({x-x'}\right)}^{2}}}{{2\tau}}}\right)
\end{equation}

with the probability of a first arrival at $\tau$ being given by:

\begin{equation}
{F_{\tau}}\left({x;x'}\right)=\frac{{\left|{x-x'}\right|}}{{\sqrt{2\pi{\tau^{3}}}}}\exp\left({-\frac{{{\left({x-x'}\right)}^{2}}}{{2\tau}}}\right)
\end{equation}

To include the mass, scale by $\tau\to\tau'\equiv m\tau$ giving:

\begin{equation}
\frac{{\partial{P_{\tau}}}}{{\partial\tau'}}=\frac{1}{{2m}}\frac{{\partial^{2}}}{{\partial{x^{2}}}}{P_{\tau'}}
\end{equation}

and:

\begin{equation}
{P_{\tau}}\left({x;x'}\right)=\frac{{\sqrt{m}}}{{\sqrt{2\pi\tau}}}\exp\left({-m\frac{{{\left({x-x'}\right)}^{2}}}{{2\tau}}}\right)\label{eq:diffusion-probability}
\end{equation}

To get the detection rate we have to handle the scaling of the time
a bit differently, since it is a probability distribution in time.
The detection rate only makes sense when integrated over $\tau$,
so the substitution implies: 

\begin{equation}
\int{d\tau{F_{\tau}}\to\int{\frac{{d\tau'}}{m}{F_{\tau'}}}}\to\int{d\tau'{D_{\tau'}}}\Rightarrow{D_{\tau}}=\frac{1}{m}{F_{\tau}}\label{eq:detection-from-first}
\end{equation}

So we have:

\begin{equation}
{D_{\tau}}\left({x-x'}\right)=\frac{{\left|{x-x'}\right|}}{\tau}\frac{{\sqrt{m}}}{{\sqrt{2\pi\tau}}}\exp\left({-m\frac{{{\left({x-x'}\right)}^{2}}}{{2\tau}}}\right)\label{eq:diffusion-detection}
\end{equation}

As a double check on this, we derive the detection rate directly from
the diffusion equation using, again, the method of images (see appendix
\ref{sec:diffusion-detection}).

\subsubsection{Wick rotation}

\label{subsec:Wick-rotation}

To go from the diffusion equation to the Schrödinger equation:

\begin{equation}
\imath\frac{\partial}{{\partial\tau}}\psi=-\frac{1}{{2{m}}}\frac{{\partial^{2}}}{{\partial{x^{2}}}}\psi
\end{equation}

we use Wick rotation:

\begin{equation}
\tau\to-\imath\tau,P\to\psi
\end{equation}

so we have the Wick-rotated kernel $K$:

\begin{equation}
{K_{\tau}}\left({x;x'}\right)=\sqrt{\frac{m}{{2\pi\imath\tau}}}\exp\left({\imath m\frac{{{\left({x-x'}\right)}^{2}}}{{2\tau}}}\right)\label{eq:free-kernel-space}
\end{equation}

and for the kernel for first arrival $F$:

\begin{equation}
{F_{\tau}}\left({x;x'}\right)=\frac{{\left|{x-x'}\right|}}{\tau}{K_{\tau}}\left({x-x'}\right)\label{eq:first-kernel-space}
\end{equation}

As a second double check, we derive $F$ directly using Laplace transforms
(see appendix \ref{sec:Alternate-derivation}).

\subsubsection{Application to a Gaussian test function}

\label{subsec:Application-to-GTF}

We start with the ``bullet'' GTF from our analysis of the Kijowski
metric (equation \ref{eq:bullet-wave-function}):

\begin{equation}
{\varphi_{0}}\left({x}\right)=\sqrt[4]{{\frac{1}{{\pi\sigma_{x}^{2}}}}}{e^{\imath{p_{0}}x-\frac{1}{{2\sigma_{x}^{2}}}{{\left({x+d}\right)}^{2}}}}
\end{equation}

with:

\begin{equation}
\left|{d}\right|\gg{\sigma_{x}}
\end{equation}

The wave function at the detector is then: 

\begin{equation}
{\varphi_{\tau}}\left(0\right)=\int\limits _{-\infty}^{0}{dx'{F_{\tau}}\left({0;x'}\right)}{\varphi_{0}}\left({x'}\right)=\int\limits _{-\infty}^{0}{dx'\frac{{\left|{0-x'}\right|}}{\tau}{K_{\tau}}\left({0;x'}\right)}{\varphi_{0}}\left({x'}\right)
\end{equation}

We can solve this explicitly in terms of error functions. However
since $\varphi_{0}$ is strongly centered on $-d$ we can use the
same trick as we did in the analysis of the Kijowski metric. We extend
the upper limit of integration to $\infty$:

\begin{equation}
{\varphi_{\tau}}\left(0\right)\approx-\frac{1}{\tau}\int\limits _{-\infty}^{\infty}{dx'x'{K_{\tau}}\left({0;x'}\right)}{\varphi_{0}}\left({x'}\right)
\end{equation}

We see by inspection we can pull down the $x'$ by taking the derivative
of $\varphi_{0}$ with respect to $p_{0}$. This gives:

\begin{equation}
{\varphi_{\tau}}\left(0\right)\approx\imath\frac{1}{{\tau}}\frac{\partial}{{\partial{p_{0}}}}\int\limits _{-\infty}^{\infty}{dx'{K_{\tau}}\left({0;x'}\right)}{\varphi_{0}}\left({x'}\right)
\end{equation}

We recognize the integral as the integral that propagates the free
wave function from $0\to\tau$:

\begin{equation}
{\varphi_{\tau}}\left(0\right)\approx\imath\frac{1}{{\tau}}\frac{\partial}{{\partial{p_{0}}}}\varphi_{\tau}^{\left({free}\right)}\left(0\right)
\end{equation}

We have:

\begin{equation}
\varphi_{\tau}^{\left({free}\right)}\left(0\right)=\sqrt[4]{{\frac{1}{{\pi\sigma_{x}^{2}}}}}\sqrt{\frac{1}{{f_{\tau}^{\left(x\right)}}}}{e^{-\frac{1}{{2\sigma_{x}^{2}f_{\tau}^{\left(x\right)}}}{{\left({d-\frac{{p_{0}}}{m}\tau}\right)}^{2}}-\imath\frac{{p_{0}^{2}}}{{2m}}\tau}}
\end{equation}

so the wave function for a first arrival at $\tau$ is:

\begin{equation}
{\varphi_{\tau}}\left(0\right)\approx\left({\imath\frac{1}{{m}}\frac{{d-{v_{0}}\tau}}{{\sigma_{x}^{2}f_{\tau}^{\left(x\right)}}}+v_{0}}\right)\varphi_{\tau}^{\left({free}\right)}\left(0\right)
\end{equation}

\subsubsection{Metrics}

\label{subsec:Metrics}

We can compute the probability density to a sufficient order by expanding
around the average time of arrival:

\begin{equation}
\bar{\tau}\equiv\frac{d}{{v_{0}}},\delta\tau\equiv\tau-\bar{\tau}
\end{equation}

The first term in the expression for the $\varphi_{\tau}$ scales
as $1/{\bar{\tau}}$; the second as a constant so at large $\tau$
we have:

\begin{equation}
{\rho_{\tau}}\left(0\right)\approx v_{0}^{2}\rho_{\tau}^{\left({free}\right)}\left(0\right)
\end{equation}

The result is the same as with Kijowski, except for the overall factor
of $v_{0}^{2}$. Since the uncertainty is produced by normalizing
this, the uncertainty in time is the same as with the bullet calculation
for Kijowski. But the factor of $v_{0}^{2}$ is troubling, particularly
when we recall we are using natural units so that at non-relativistic
speeds this is hard to distinguish from zero.

The immediate problem is our somewhat thin-skinned approximation.
The random walk can only penetrate to position $m=0$ in the grid
before it is absorbed. And since the probability density is computed
as the square, the result is doubly small. 

Given the taking of the limit, it is perhaps more surprising that
we get a finite detection rate at all. A realistic approximation would
have the paths penetrating some finite distance into the detector,
with the absorption going perhaps with some characteristic length
-- as with the Marchewka and Schuss approach.

In return for a \emph{more} detailed examination of the boundary we
have gotten a \emph{less} physically realistic estimate. But the real
problem with all three of the approaches we have looked at is their
assumption of a crisp boundary between quantum and classical realms.

\subsection{Implications for time-of-arrival measurements}
\begin{quote}
“What is it that we human beings ultimately depend on? We depend on
our words. We are suspended in language. Our task is to communicate
experience and ideas to others. We must strive continually to extend
the scope of our description, but in such a way that our messages
do not thereby lose their objective or unambiguous character.'' --
Niels Bohr as cited in \cite{Petersen:1963aa}
\end{quote}
In reality there can be no crisp boundary between quantum and classical
realms. Consider the lifespan of an atom built on classical principles,
as in say Bohr's planetary model of the atom. A classical electron
in orbit around the nucleus undergoes centripetal acceleration from
the nucleus; therefore emits Larmor radiation; therefore decays in
towards the nucleus. The estimated lifespan is of order $10^{-11}$
seconds \cite{Olsen:2017aa}. 

What keeps this from happening in the case of a quantum atom is the
uncertainty principle: as the electron spirals into the nucleus its
uncertainty in position is reduced so its uncertainty in momentum
is increased. A larger $\Delta p$ means a larger kinetic energy.
In fact the ground state energy can be estimated as the minimum total
energy -- kinetic plus potential -- consistent with the uncertainty
principle.

Since everything material is built of atoms and since there are no
classical atoms there is no classical realm. And therefore it is incorrect
to speak of a transition to a classical end state. No such state exists.

However classical mechanics does work very well in practice. So what
we have is a problem in description. Which parts of our experimental
setup -- particle, source, detector, observers -- can be described
to acceptable accuracy using classical language; where must we use
a fully quantum description?

This approach places us, with Bohr \cite{Camilleri:2015aa}, firmly
on the epistemological side of the measurement debate. One possible
starting point is the program of decoherence \cite{Zurek:2002ky,Omnes:1994ao,Giulini:1996vp,Zeh:1996aa,Zeh:2000sj,Kim:2002aa,Joos:2003gf,Schlosshauer:2007rr}.
The key question for decoherence is not when does a system go from
quantum to classical but rather when does a classical description
become possible. Another is Quantum Trajectory Theory (QTT) \cite{Carmichael:1991aa,Dalibard:1992aa,Dum:1992aa,Hegerfeldt:1991aa}.
QTT, as its name suggests, is a particularly good fit with the path
integral approach, the formalism we are primarily focused on here.

For the moment we will explore a more informal approach, taking advantage
of the conceptual simplicity of the path integral approach. We can
describe the evolution of a quantum particle in terms of the various
paths it takes. In the case of the problem at hand, we visualize the
paths as those that enter the detector or do not, that are reflected
or are not, that leave and do not return (backflow) or return again
and so on. Ultimately however each of the paths will escape permanently
or else enter an entropy sink of some kind, a structure that is sufficiently
macroscopic that it is no longer possible to analyze the process in
terms of the individual paths. Its effects will live on, perhaps in
a photo-electric cascade or a change of temperature. But the specific
path will be lost like a spy in a crowd.

The entropy sink is the one essential part needed for a measurement
to take place. In a quantum mechanical system information, per the
no-cloning theorem, is neither created nor destroyed. But the heart
of a measurement is the reduction of the complexity of the system
under examination to a few numbers. Information is necessarily lost
in doing this. In general it is fairly clear where this happens; typically
in crossing the boundary from the microscopic to the macroscopic scale.

\begin{figure}
\includegraphics[scale=0.67]{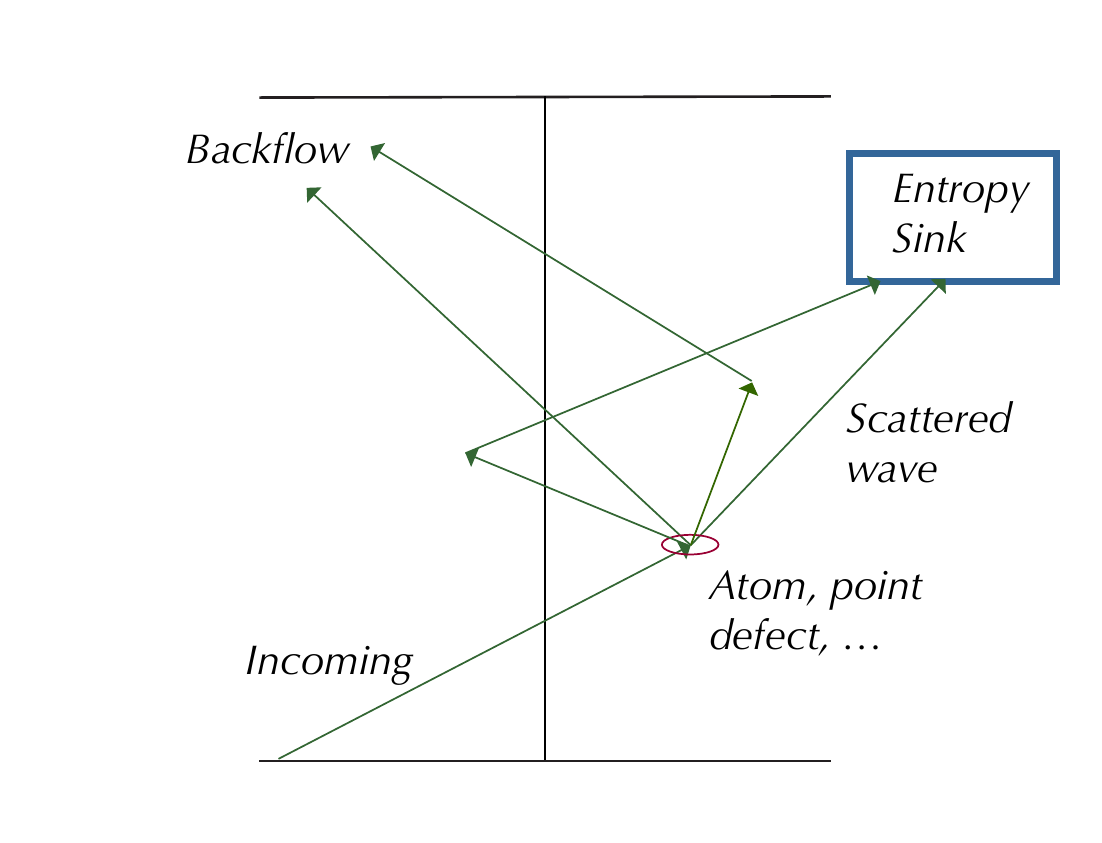}

\caption{Global character of the measurement problem}
\end{figure}

Typical paths may cross the boundary multiple times, ultimately either
being absorbed and registered (measured), lost in some way (detector
inefficiency), or reflected (backflow). There will be some time delay
associated with all of these cases\footnote{A typical representation could be done in terms of a Laplace transform
$\mathcal{D}\left(s\right)$ of the detection amplitude plus a Laplace
transform $\mathcal{R}\left(s\right)$ of the reflection amplitude.
The Marchewka and Schuss fudge parameter may be thought of as a useful
first approximation of $\mathcal{D}$.}. 

This is not too far from conventional practice. A macroscopic detector
functions as the entropy sink, its efficiency is usually known, and
reflections are possible but minimized. In many cases the response
time associated with the process of detection may not require much
attention -- we may only care that there was a detection. Our case
is a bit trickier than that. We have to care about the time delay
associated with a detection, especially if the time delay introduces
some uncertainty in time in its own right.

This picture is enough to show that there can not be, in general,
a local time-of-arrival operator. Suppose there is such. Consider
the paths leaving it on the right, continuing to an entropic sink
of some kind. If we arrange a quantum mirror of some kind (perhaps
just a regular mirror), then some of the paths will return to their
source. The detection rate will be correspondingly reduced. But this
can only be predicted using knowledge of the global situation, of
the interposed mirror. An operator local to the boundary cannot know
about the mirror and therefore cannot give a correct prediction of
the time-of-arrival distribution. Therefore the time-of-arrival cannot
in general be given correctly by a purely local operator.

Since the fundamental laws are those of quantum mechanics, the analysis
must be carried out at a quantum mechanical level -- except for those
parts where we can show the classical approximation suffices, usually
the end states. In fact, there is no guarantee that even the end states
can be adequately described in classical terms; while there are no
cases -- so far -- of a perdurable and macroscopic quantum system\footnote{But see the macroscopic wave functions in \cite{Kovachy:2015aa,Asenbaum:2017aa}.},
there is nothing to rule that out.

\section{Effects of dispersion in time on time-of-arrival measurements}

\label{sec:TQM}
\begin{quote}
``Clearly,'' the Time Traveller proceeded, ``any real body must
have extension in four directions: it must have Length, Breadth, Thickness,
and --- Duration. But through a natural infirmity of the flesh, which
I will explain to you in a moment, we incline to overlook this fact.
There are really four dimensions, three which we call the three planes
of Space, and a fourth, Time. There is, however, a tendency to draw
an unreal distinction between the former three dimensions and the
latter, because it happens that our consciousness moves intermittently
in one direction along the latter from the beginning to the end of
our lives.'' -- H. G. Wells \cite{Wells:1935dc}
\end{quote}
We now return to our original problem, first estimating the dispersion
in time-of-arrival without dispersion in time, then with.

\subsection{Time-of-arrival measurements without dispersion in time}

\label{subsec:TOA-in-SQM}

Our efforts to correctly define a time of arrival operator have led
to the conclusion that the time-of--arrival -- like measurement
in general -- is a system level property; it is not something that
can be correctly described by a local operator.

We have therefore to a certain extent painted ourselves into a corner.
We would like to compute the results for the general case, but we
have shown there are only specific cases. We will deal with this by
trading precision for generality.

We will first assume a perfect detector, then justify the assumption.

\subsubsection{Probability current in SQM}

\label{subsec:Probability-current-sqm}

We will first compute the detection rate by using the probability
current per Marchewka and Schuss \cite{Marchewka:2000ys}. We start
with the Schrödinger equation:

\begin{equation}
\imath\frac{\partial}{{\partial\tau}}{\psi_{\tau}}\left(x\right)=-\frac{1}{{2m}}\frac{{\partial^{2}}}{{\partial{x^{2}}}}{\psi_{\tau}}\left(x\right)\label{eq:schrodinger-equation}
\end{equation}

We will assume we have in hand a wave function that includes the interaction
with the detector. We define the probability of detection at clock
time $\tau$ as $D_{\tau}$.

The total probability of the wave function to be found to the left
of the detector at time $\tau$ is the integral of the probability
density from $-\infty\to0$. The total probability to have been detected
up to time $\tau$ is the integral of $D_{\tau}$ from $0\to\tau$.
Therefore by conservation of probability we have:

\begin{equation}
1=\int\limits _{-\infty}^{0}{dx}\psi_{\tau}^{*}\left(x\right){\psi_{\tau}}\left(x\right)+\int\limits _{0}^{\tau}{d\tau'{D_{\tau'}}}
\end{equation}

We take the derivative with respect to $\tau$ to get:

\begin{equation}
D_{\tau}=-\int\limits _{-\infty}^{0}{dx\frac{{\partial\psi_{\tau}^{*}\left(x\right)}}{{\partial\tau}}{\psi_{\tau}}\left(x\right)+\psi_{\tau}^{*}\left(x\right)\frac{{\partial{\psi_{\tau}}\left(x\right)}}{{\partial\tau}}}
\end{equation}

We use the Schrödinger equation:

\begin{equation}
{D_{\tau}}=\frac{\imath}{{2m}}\int\limits _{-\infty}^{0}{dx\frac{{{\partial^{2}}\psi_{\tau}^{*}\left(x\right)}}{{\partial{x^{2}}}}{\psi_{\tau}}\left(x\right)-\psi_{\tau}^{*}\left(x\right)\frac{{{\partial^{2}}{\psi_{\tau}}\left(x\right)}}{{\partial{x^{2}}}}}\label{eq:integral-by-parts}
\end{equation}

And integrate by parts to get the equation for the detection rate:

\begin{equation}
{D_{\tau}}={\left.{\frac{\imath}{{2m}}\left({\frac{{\partial\psi_{\tau}^{*}\left(x\right)}}{{\partial x}}{\psi_{\tau}}\left(x\right)-\psi_{\tau}^{*}\left(x\right)\frac{{\partial{\psi_{\tau}}\left(x\right)}}{{\partial x}}}\right)}\right|_{x=0}}
\end{equation}

We recognize the expression on the right as the probability current:

\begin{equation}
{J_{\tau}}\left(x\right)=-\frac{\imath}{{2}}\left({\psi_{\tau}^{*}\left(x\right)\frac{{\partial{\psi_{\tau}}\left(x\right)}}{{\partial x}}-{\psi_{\tau}}\left(x\right)\frac{{\partial\psi_{\tau}^{*}\left(x\right)}}{{\partial x}}}\right)
\end{equation}

or:

\begin{equation}
{J_{\tau}}=\psi_{\tau}^{*}\frac{p}{{2m}}{\psi_{\tau}}-{\psi_{\tau}}\frac{p}{{2m}}\psi_{\tau}^{*}\label{eq:MS-probability-current}
\end{equation}

so:

\begin{equation}
D_{\tau}=J_{\tau}
\end{equation}

Note that $D_{\tau}$ can, in the general case, be negative. For instance,
if the detector includes -- per the previous section -- a mirror
of some kind there may be backflow, probability current going from
right to left.

Ironically enough, this is in fact the first metric that Kijowski
considered in his paper, only to reject it because it violated the
classical condition \ref{classical-condition}. 

\subsubsection{Black box detector}

\label{subsec:The-clock-in-a-box-in-reverse}

Given the wave function we can compute the detection rate using the
probability current. The real problem is to compute the wave function
in the first place, given that this must in principle include the
interaction of the particle with the detector. To treat the general
case, we would like a general detector, one where the details of the
interaction do not matter. We are looking for something which is a
perfect black.

We can use Einstein's clock-in-a-box, but this time run it in reverse:
have the detector be a box that has a small window open for a fraction
of a second then check how many particles are in the box afterward
by weighing the box. The interior walls of the box provide the necessary
entropy sink. We will refer to this as a black box detector \footnote{A ``perfect black'' is apparently not quite as theoretical a concept
as it sounds: the commercial material Vantablack can absorb 99.965\%
of incoming light \cite{Jackson:2010aa}. And Vantablack even uses
a similar mechanism: incoming light is trapped in vertically aligned
nanotube arrays.}.

In fact, this is a reasonable model of the elementary treatment of
detection. When, for instance, we compute the rate of scattering of
particles into a bit of solid angle, we assume that the particles
will be absorbed with some efficiency but we do not generally worry
about subtleties as time delay till a detection is registered or backflow.

\subsubsection{Application to a Gaussian test function}

We will take a GTF at the plane as above:

\begin{equation}
{\varphi_{\tau}}\left(x\right)=\sqrt[4]{{\frac{1}{{\pi\sigma_{x}^{2}}}}}\sqrt{\frac{1}{{f_{\tau}^{\left(x\right)}}}}{e^{\imath{p_{0}}x-\frac{1}{{2\sigma_{x}^{2}f_{\tau}^{\left(x\right)}}}{{\left({x+d-\frac{{p_{0}}}{m}{\tau}}\right)}^{2}}-\imath\frac{{p_{0}^{2}}}{{2m}}{\tau}}}\label{eq:sqm-wf-at-detector}
\end{equation}

An elementary application of the probability current gives the detection
rate as the velocity times the probability density:

\begin{equation}
{D_{\tau}}=\frac{{p_{0}}}{m}{\rho_{\tau}}\left(0\right)\label{eq:detection-sqm}
\end{equation}

This is the same as we had for the Kijowski bullet (equation \ref{eq:detection-rate}).
And it is, as noted there, also the classical rate for detection of
a probability distribution impacting a wall. 

Our use of the Einstein box in reverse is intended to provide a lowest
common denominator for quantum measurements; it is not surprising
that the lowest common denominator for quantum measurements should
be the corresponding classical distribution. 

We have the uncertainty in time-of-arrival as before (equation \ref{eq:kijowski-bullet-uncertainty}):

\begin{equation}
\Delta\tau=\frac{1}{{\sqrt{2}}}{{\bar{\sigma}}_{\tau}}=\frac{1}{{\sqrt{2}}}\frac{1}{{m{v_{0}}{\sigma_{x}}}}\bar{\tau}\label{eq:sqm-uncertainty}
\end{equation}

Since we have $\bar{p}=m{v_{0}}$ and -- for a minimum uncertainty
wave function -- ${\sigma_{p}}=\frac{1}{{\sigma_{x}}}$ we can also
write:

\begin{equation}
\frac{{\sigma_{\tau}}}{{\bar{\tau}}}=\frac{{{\hat{\sigma}}_{p}}}{{\bar{p}}}
\end{equation}

so the dispersion in momentum and in clock time are closely related,
as one might expect.

While the results are the same as with the Kijowski metric (for bullet
wave functions), the advantage is that we now have a clearer sense
of what is actually going on and therefore what corrections we might
have to make in practice. These may include (but will probably not
be limited to):
\begin{enumerate}
\item Depletion, e.g. inefficiencies of the detector,
\item Backflow, e.g. a reflection coefficient,
\item Delay in the detection (not normally important, but significant here),
\item Self-interference of the wave function with itself,
\item Edge effects -- one can imagine the incoming part of the wave function
skimming along the edge of the detector for instance,
\item And of course the increasingly complex effects that result from increasingly
complex models of the wave function-detector interaction.
\end{enumerate}

\subsection{Time-of-arrival measurements with dispersion in time}

\subsubsection{Quantum mechanics with uncertainty in time}

\label{subsec:tqm-recap-uncert}

\begin{figure}[h]
\begin{centering}
\includegraphics[scale=0.33]{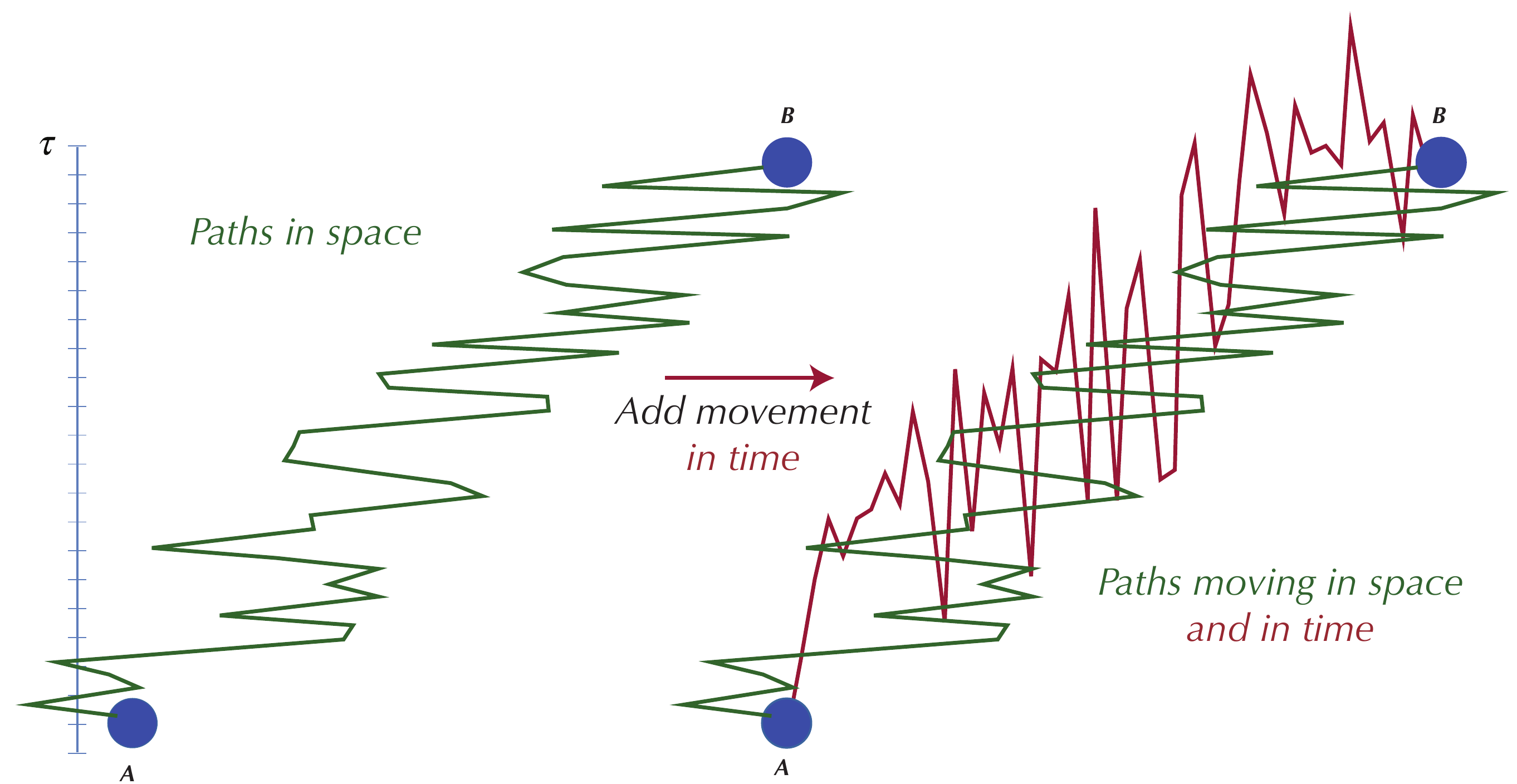}
\par\end{centering}
\caption{Paths in three and in four dimensions\label{fig:paths-in-space-and-time}}
\end{figure}

Path integrals provide a natural way to extend quantum mechanics to
include time in a fully covariant way. In \cite{Ashmead_2019} we
did this by first extending the usual three-dimensional paths to four
dimensions. The rest of the treatment followed from this single assumption
together with the twin requirements of covariance and of consistency
with known results. But as one might expect, there are a fair number
of questions to be addressed along the way -- making the full paper
both rather long and rather technical. To make this treatment self-contained
we summarize the key points here.

\paragraph{Feynman path integrals in four dimensions}

\label{subsec:paths-in-time}

We start with clock time, defined operationally as what the laboratory
clock measures. This is the parameter $\tau$ in the Schrödinger equation,
Klein-Gordon equation, path integral sums, and so on. 

The normal three dimensional paths are parameterized by clock time;
we generalize them to four dimensional paths, introducing the coordinate
time $t$ to do so:

\begin{equation}
{\bar{\pi}_{\tau}}\left({x,y,z}\right)\to{\pi_{\tau}}\left({t,x,y,z}\right)
\end{equation}

It can be helpful to think of coordinate time as another space dimension.
We will review the relationship between clock time and coordinate
time at the end, once the formalism has been laid out. (This is a
case where it is helpful to let the formalism precede the intuition.)

In path integrals we get the amplitude to go from one point to another
by summing over all intervening paths, weighing each path by the corresponding
action. The path integral with four dimensional paths is formally
identical to the path integral with three dimensions, except that
the paths take all possible routes in four rather than three dimensions.
We define the kernel as:

\begin{equation}
{K_{\tau}}\left({x'';x'}\right)=\mathop{\lim}\limits _{N\to\infty}\int\mathcal{D}\pi{e^{\imath\int\limits _{0}^{\tau}{d\tau'\mathcal{L}\left[{\pi,\dot{\pi}}\right]}}}
\end{equation}

We use a Lagrangian which works for both the three and four dimensional
cases (see \cite{Goldstein:1980ce}):

\begin{equation}
\mathcal{L}\left({{x^{\mu}},{{\dot{x}}^{\mu}}}\right)=-\frac{1}{2}m{{\dot{x}}^{\mu}}{{\dot{x}}_{\mu}}-q{{\dot{x}}^{\mu}}{A_{\mu}}\left(x\right)-\frac{m}{2}\label{eq:lagragian-4d}
\end{equation}

As usual with path integrals, to actually do the integrals we discretize
clock time $\tau=\epsilon n$ and replace the integral over clock
time with a sum over steps:

\[
\int\limits _{0}^{\tau}{d\tau'\mathcal{L}}\to\sum\limits _{j=0}^{n}{\mathcal{L}_{j}}
\]

with:

\begin{equation}
\mathcal{L}_{j}\equiv-m\frac{\left(x_{j}-x_{j-1}\right)^{2}}{2\varepsilon^{2}}-q\frac{x_{j}-x_{j-1}}{\epsilon}\frac{A\left(x_{j}\right)+A\left(x_{j-1}\right)}{2}-\frac{m}{2}\label{eq:lagrangian}
\end{equation}

And measure:
\begin{equation}
\mathcal{D}\pi\equiv\left(-\imath\frac{m^{2}}{4\pi^{2}\varepsilon^{2}}\right)^{N}\prod\limits _{n=1}^{N-1}d^{4}x_{n}
\end{equation}

This gives us the ability to compute the amplitude to get from one
point to another in a way that includes paths that vary in time as
well as space.

\paragraph{Schrödinger equation in four dimensions}

Usually we derive the path integral expressions from the Schrödinger
equation. But here we derive the four dimensional Schrödinger equation
from the short time limit of the path integral kernel, running the
usual derivation in reverse and with one extra dimension. We get:

\begin{equation}
\imath\frac{\partial\psi_{\tau}}{\partial\tau}=-\frac{1}{{2m}}\left({\left({{p_{\mu}}-q{A_{\mu}}}\right)\left({{p^{\mu}}-q{A^{\mu}}}\right)-m^{2}}\right){\psi_{\tau}}\label{eq:schroedinger-equation-4d}
\end{equation}

Note that here $p$ is an operator, $A$ is an external field, and
$m$ is the constant operator\footnote{In the extension to QED, $A$ becomes an operator as well.}.
This equation goes back to Stueckelberg \cite{Stueckelberg:1941aa,Stueckelberg:1941la}
with further development by Feynman, Fanchi, Land, and Horwitz \cite{Feynman:1948,Fanchi:1993ab,Land:1996aj,Horwitz:2015jk}.

We need only the free Schrödinger equation here. With the vector potential
$A$ set to zero we have:

\begin{equation}
\imath\frac{\partial}{{\partial\tau}}{\psi_{\tau}}=-\frac{{{E^{2}}-{\overrightarrow{p}^{2}}-m^{2}}}{{2m}}{\psi_{\tau}}
\end{equation}

Or spelled out:

\begin{equation}
\imath\frac{\partial}{{\partial\tau}}{\psi_{\tau}}\left({t,\vec{x}}\right)=\frac{1}{{2m}}\left({\frac{{\partial^{2}}}{{\partial{t^{2}}}}-{\nabla^{2}}-{m^{2}}}\right){\psi_{\tau}}\left({t,\vec{x}}\right)\label{eq:schrodinger-equation-4d-free}
\end{equation}

If the left hand side were zero, the right hand side would be the
Klein-Gordon equation (with $t\to\tau$). Over longer times -- femtoseconds
or more -- the left side will in general average to zero, giving
the Klein-Gordon equation as the long time limit. More formally we
expect that if we average over times $\Delta\tau$ of femtoseconds
or greater, we will have:

\begin{equation}
\int\limits _{\tau}^{\tau+\Delta\tau}{d\tau'\int{{d^{4}}x}}\psi_{\tau}^{*}\left(x\right)\frac{{{E^{2}}-{{\vec{p}}^{2}}-{m^{2}}}}{{2m}}{\psi_{\tau}}\left(x\right)\approx0
\end{equation}

But at short times -- attoseconds -- we should see effects associated
with uncertainty in time. This is in general how we get from a fully
four dimensional approach to the usual three dimensional approach;
the fluctuations in coordinate time tend to average out over femtosecond
and longer time scales. But at shorter times -- attoseconds -- we
will see the effects of uncertainty in time. 

It is much the same way that in SQM quantum effects in space average
out over longer time and distance scales to give classical mechanics.
TQM is to SQM -- with respect to time -- as SQM is to classical
mechanics with respect to the three space dimensions.

\paragraph{Wave functions in coordinate time}

We need an estimate of the wave function at source, its evolution
in clock time, and the rules for detection -- the birth, life, and
death of the wave function if you will.

\subparagraph{Initial wave function}

\label{par:Initial-wave-function}

We start with the free Klein-Gordon equation and use the method of
separation of variables to break the wave function out into a space
and a time part:

\begin{equation}
\varphi_{0}\left(t,x\right)=\tilde{\varphi}_{0}\left(t\right)\bar{\varphi}_{0}\left(x\right)
\end{equation}

Or in energy/momentum space:

\begin{equation}
\hat{\varphi}_{0}\left(E,p\right)=\hat{\tilde{\varphi}}_{0}\left(E\right)\hat{\bar{\varphi}}_{0}\left(p\right)
\end{equation}

We assume we already have the space part by standard methods. For
instance, if this is a GTF in momentum space it will look like:

\begin{equation}
{{\hat{\bar{\varphi}}}_{\tau}}\left(p\right)=\sqrt[4]{{\frac{1}{{\pi\sigma_{p}^{2}}}}}{e^{-\imath p{x_{0}}-\frac{{{\left({p-{p_{0}}}\right)}^{2}}}{{2\sigma_{p}^{2}}}-\imath\frac{{p^{2}}}{{2m}}\tau}}
\end{equation}

This plus the Klein-Gordon equation give us constraints on the time
part:

\begin{equation}
\begin{array}{l}
\left\langle 1\right\rangle =1\\
\bar{E}\equiv\left\langle E\right\rangle =\sqrt{m^{2}+\left\langle \vec{p}\right\rangle ^{2}}\\
\left\langle E^{2}\right\rangle =\left\langle m^{2}+\vec{p}^{2}\right\rangle =m^{2}+\left\langle \vec{p}^{2}\right\rangle 
\end{array}\label{eq:constraints-1}
\end{equation}

To get a robust estimate of the shape of the time part we assume it
is the maximum entropy solution that satisfies the constraints. We
used the method of Lagrange multipliers to find this getting:

\begin{equation}
\hat{\tilde{\varphi}}_{0}\left(E\right)=\sqrt[4]{\frac{1}{\pi\sigma_{E}^{2}}}e^{\imath\left(E-E_{0}\right)t_{0}-\frac{\left(E-E_{0}\right)^{2}}{2\sigma_{E}^{2}}}
\end{equation}

with values: 
\begin{equation}
\sigma_{E}^{2}=\sigma_{p}^{2}
\end{equation}
and: 
\begin{equation}
E_{0}=\sqrt{m^{2}+\bar{p}^{2}}
\end{equation}

To get the starting wave function in time we take the inverse Fourier
transform: 
\begin{equation}
\tilde{\varphi}_{0}\left(t\right)=\sqrt[4]{\frac{1}{\pi\sigma_{t}^{2}}}e^{-\imath E_{0}\left(t-t_{0}\right)-\frac{t^{2}}{2\sigma_{t}^{2}}}\label{eq:toa-tqm-recap-initial-wf-time}
\end{equation}
\begin{equation}
\sigma_{t}^{2}=\frac{1}{\sigma_{E}^{2}}
\end{equation}

The value of the $\sigma_{t}$ will normally be of order attoseconds.
We set $t_{0}=0$ as the overall phase is already supplied by the
space/momentum part.

Since estimates from maximum entropy tend to be robust we expect this
approach will give estimates which are order-of-magnitude correct.

\subparagraph{Propagation of a wave function}

\label{par:Propagation-of-wf}

The TQM form of the Klein-Gordon equation is formally identical to
the non-relativistic Schrödinger equation with the additional coordinate
time term. In momentum space one can use this insight to write the
kernel out by inspection:

\begin{equation}
{{\hat{K}}_{\tau}}\left({p'';p'}\right)={\delta^{\left(4\right)}}\left({p''-p'}\right)\exp\left({\imath\frac{{{{p'}^{2}}-{m^{2}}}}{{2m}}\tau}\right)\theta\left(\tau\right)
\end{equation}

The coordinate space form is:

\begin{equation}
{K_{\tau}}\left({x'';x'}\right)=-\imath\frac{{m^{2}}}{{4{\pi^{2}}{\tau^{2}}}}{e^{-\imath m\frac{{{\left({x''-x'}\right)}^{2}}}{{2\tau}}-\imath\frac{m}{2}\tau}}\theta\left(\tau\right)
\end{equation}

In both cases we have the product of a coordinate time part by the
familiar space part. Spelling this out in coordinate space\footnote{We are using an over-tilde and over-bar to distinguish between the
coordinate time and the space parts, when this is useful.}:

\begin{equation}
\begin{gathered}{K_{\tau}}\left({x'';x'}\right)={{\tilde{K}}_{\tau}}\left({{t^{\prime\prime}};{t^{\prime}}}\right){{\bar{K}}_{\tau}}\left({{{\vec{x}}^{\prime\prime}};{{\vec{x}}^{\prime}}}\right)\exp\left({-\imath\frac{m}{2}\tau}\right)\theta\left(\tau\right)\hfill\\
{{\tilde{K}}_{\tau}}\left({{t^{\prime\prime}};{t^{\prime}}}\right)=\sqrt{\frac{{\imath m}}{{2\pi\tau}}}\exp\left({-\imath m\frac{{{\left({{t^{\prime\prime}}-{t^{\prime}}}\right)}^{2}}}{{2\tau}}}\right)\hfill\\
{{\bar{K}}_{\tau}}\left({{{\vec{x}}^{\prime\prime}};{{\vec{x}}^{\prime}}}\right)={\sqrt{-\frac{\imath m}{{2\pi\tau}}}^{3}}\exp\left({\imath m\frac{{{\left({\vec{x}''-\vec{x}'}\right)}^{2}}}{{2\tau}}}\right)\hfill
\end{gathered}
\label{eq:tqm-kernel}
\end{equation}

where the space part kernel is merely the familiar non-relativistic
kernel (e.g. Merzbacher \cite{Merzbacher:1998tc}). The additional
factor of $\exp\left({-\imath\frac{m}{2}\tau}\right)$ only contributes
an overall phase which plays no role in single particle calculations.

If the initial wave function can be written as the product of a coordinate
time and a three-space part, then the propagation of the three-space
part is the same as in non-relativistic quantum mechanics. In general,
if the free wave function starts as a direct product in time and space
it stays that way.

Of particular interest here is the behavior of a GTF in time. If at
clock time zero it is given by:

\begin{equation}
\tilde{\varphi}_{0}\left(t\right)\equiv\sqrt[4]{\frac{1}{\pi\sigma_{t}^{2}}}e^{-\imath E_{0}\left(t-t_{0}\right)-\frac{\left(t-t_{0}\right)^{2}}{2\sigma_{t}^{2}}}\label{eq:wf-time-0}
\end{equation}

then as a function of clock time it is:

\begin{equation}
\tilde{\varphi}_{\tau}\left(t\right)=\sqrt[4]{\frac{1}{\pi\sigma_{t}^{2}}}\sqrt{\frac{1}{f_{\tau}^{\left(t\right)}}}e^{-\imath E_{0}t-\frac{1}{2\sigma_{t}^{2}f_{\tau}^{\left(t\right)}}\left(t-t_{0}-\frac{E_{0}}{m}\tau\right)^{2}+\imath\frac{E_{0}^{2}}{2m}\tau}\label{eq:wf-time-tau}
\end{equation}
with dispersion factor $f_{\tau}^{\left(t\right)}\equiv1-\imath\frac{\tau}{m\sigma_{t}^{2}}$
and with expectation, probability density, and uncertainty: 
\begin{equation}
\begin{array}{l}
\left\langle {t_{\tau}}\right\rangle ={t_{0}}+\frac{E}{m}\tau={t_{0}}+\gamma\tau\\
{{\tilde{\rho}}_{\tau}}\left(t\right)=\sqrt{\frac{1}{{\pi\sigma_{t}^{2}}\left({1+\frac{{\tau^{2}}}{{{m^{2}}\sigma_{t}^{4}}}}\right)}}\exp\left({-\frac{{{\left({t-\left\langle {t_{\tau}}\right\rangle }\right)}^{2}}}{{\sigma_{t}^{2}\left({1+\frac{{\tau^{2}}}{{{m^{2}}\sigma_{t}^{4}}}}\right)}}}\right)\\
{\left({\Delta t}\right)^{2}}\equiv\left\langle t^{2}\right\rangle -{\left\langle t\right\rangle ^{2}}=\frac{{\sigma_{t}^{2}}}{2}\left|{1+\frac{{\tau^{2}}}{{{m^{2}}\sigma_{t}^{4}}}}\right|
\end{array}
\end{equation}

The behavior of the time part is given by replacing $x\to t$ and
taking the complex conjugate. As noted, coordinate time behaves like
a fourth space dimension, albeit one that enters with the sign of
$\imath$ flipped. 

\subparagraph{Detection of a wave function}

\label{subsec:Birth-and-death}

To complete the birth, life, death cycle we look at the detection
of a wave function. This is obviously the place where things can get
very tricky. However the general rule that coordinate time acts as
a fourth space coordinate eliminates much of the ambiguity. If we
were doing a measurement along the $y$ dimension and we registered
a click in a detector located at $y_{detector}\pm\frac{\Delta y}{2}$
we would take this as meaning that we had measured the particle as
being at position $y_{detector}$ with uncertainty $\Delta y$. 

Because of our requirement of the strictest possible correspondence
between coordinate time and the space dimensions, the same rule necessarily
applies in coordinate time. If we have a clock in the detector with
time resolution $\Delta\tau$, then a click in the detector implies
a measurement of the particle at coordinate time $t=\tau$ with uncertainty
$\Delta\tau.$ 

So the detector does an implicit conversion from coordinate time to
clock time, because we have assumed we know the position of the detector
in clock time. How do we justify this assumption?

The justification is that we take the clock time of the detector as
being itself the average of the coordinate time of the detector:

\begin{equation}
{\tau_{Detector}}\equiv\left\langle {t_{Detector}}\right\rangle 
\end{equation}

The detector will in general be made up of a large number of individual
particles, each with a four-position in coordinate time and the three
space dimensions. While in TQM, the individual particles may be a
bit in the future or past, the average will behave as a classical
variable, specifically as the clock time. This drops out of the path
integral formalism, which supports an Ehrenfest's theorem in time
(again in \cite{Ashmead_2019}). If we take the coordinate time of
a macroscopic object as the sum of the coordinate times of its component
particles, the principle is macroscopically stronger. The motion in
time of a macroscopic object will no more display quantum fluctuations
in time than its motion in space displays quantum fluctuations in
the $x,y,z$ directions.

This explains the differences in the properties of the clock time
and coordinate time. The clock time is something we think of as only
going forward; its corresponding energy operator is bounded below,
usually by zero. But if it is really an expectation of the coordinate
time of a macroscopic object, then it is not that the clock time cannot
go backward, it is that it is statistically extremely unlikely to
do so. The principle is the same as the argument that while the gas
molecules in a room could suddenly pile up in one half of a room and
leave the other half a vacuum, they are statistically extremely unlikely
to do so.

Further the conjugate variable for coordinate time ${E^{\left({op}\right)}}\equiv\imath\frac{\partial}{{\partial t}}$
is not subject to Pauli's theorem \cite{Pashby:2014wu}. The values
of $E$ -- the momentum conjugate to the coordinate time $t$ --
 can go negative. If we look at a non-relativistic GTF:

\begin{equation}
{{\hat{\tilde{\varphi}}}_{0}}\left(E\right)=\sqrt[4]{{\frac{1}{{\pi\sigma_{E}^{2}}}}}{e^{\imath E{t_{0}}-\frac{{{\left({E-{E_{0}}}\right)}^{2}}}{{2\sigma_{E}^{2}}}}},{\sigma_{E}}=\frac{1}{{\sigma_{t}}},{E_{0}}\approx m+\frac{{{\vec{p}}^{2}}}{{2m}}
\end{equation}

the value of $m$ for an electron is about 500,000 eV, while for $\sigma_{t}$
of order attoseconds the $\sigma_{E}$ will be of order 6000 eV. Therefore
the negative energy part is about $500000/6000$ or $83$ standard
deviations away from the average. And therefore the likelihood of
detecting a negative energy part of the wave function is zero to an
excellent approximation. But not \emph{exactly} zero.

So, clock time is the classical variable corresponding to the fully
quantum mechanical coordinate time. They are two perspectives on the
same universe.

\subsubsection{Probability current in TQM}

\label{subsec:Probability-current-tqm}

We now compute the probability current in TQM, working in parallel
to the Marchewka and Schuss derivation for SQM covered above (and
with similar caveats about its applicability). 

By conservation of probability the sum of the detections as of clock
time $\tau$ plus the total probability remaining at $\tau$ must
equal one:

\begin{equation}
1=\int\limits _{0}^{\tau}{d\tau'\int\limits _{-\infty}^{\infty}{dt{D_{\tau'}}\left(t\right)}}+\int\limits _{-\infty}^{0}{dx}\int\limits _{-\infty}^{\infty}{dt}\psi_{\tau}^{*}\left({t,x}\right){\psi_{\tau}}\left({t,x}\right)
\end{equation}

Again we take the derivative with respect to clock time to get:

\begin{equation}
\int\limits _{-\infty}^{\infty}{dt{D_{\tau}}\left(t\right)}=-\int\limits _{-\infty}^{0}{dx}\int\limits _{-\infty}^{\infty}{dt}\left({\dot{\psi}_{\tau}^{*}\left({t,x}\right){\psi_{\tau}}\left({t,x}\right)+\psi_{\tau}^{*}\left({t,x}\right){{\dot{\psi}}_{\tau}}\left({t,x}\right)}\right)
\end{equation}

From the Schrödinger equation:

\begin{equation}
\begin{gathered}{{\dot{\psi}}_{\tau}}=-\imath\frac{1}{{2m}}\left({\frac{{\partial^{2}}}{{\partial{t^{2}}}}-\frac{{\partial^{2}}}{{\partial{x^{2}}}}}\right){\psi_{\tau}}\hfill\\
\dot{\psi}_{\tau}^{*}=\imath\frac{1}{{2m}}\left({\frac{{\partial^{2}}}{{\partial{t^{2}}}}-\frac{{\partial^{2}}}{{\partial{x^{2}}}}}\right)\psi_{\tau}^{*}\hfill
\end{gathered}
\end{equation}

We use integration by parts in coordinate time to show the terms in
the second derivative of coordinate time cancel:

\begin{equation}
\imath\frac{1}{{2m}}\left({\frac{{\partial^{2}}}{{\partial{t^{2}}}}\psi_{\tau}^{*}}\right){\psi_{\tau}}-\imath\frac{1}{{2m}}\psi_{\tau}^{*}\frac{{\partial^{2}}}{{\partial{t^{2}}}}{\psi_{\tau}}\to-\imath\frac{1}{{2m}}\left({\frac{\partial}{{\partial t}}\psi_{\tau}^{*}}\right)\frac{\partial}{{\partial t}}{\psi_{\tau}}+\imath\frac{1}{{2m}}\left({\frac{\partial}{{\partial t}}\psi_{\tau}^{*}}\right)\frac{\partial}{{\partial t}}{\psi_{\tau}}=0
\end{equation}

leaving:

\begin{equation}
\mathop{\smallint}\limits _{-\infty}^{\infty}dt{D_{\tau}}\left(t\right)=\frac{\imath}{{2m}}\mathop{\smallint}\limits _{-\infty}^{\infty}dt\mathop{\smallint}\limits _{-\infty}^{0}dx\left({\left({\frac{{{\partial^{2}}\psi_{\tau}^{*}\left({t,x}\right)}}{{\partial{x^{2}}}}}\right){\psi_{\tau}}\left({t,x}\right)-\psi_{\tau}^{*}\left({t,x}\right)\left({\frac{{{\partial^{2}}{\psi_{\tau}}\left({t,x}\right)}}{{\partial{x^{2}}}}}\right)}\right)\label{eq:probability-current-4d}
\end{equation}

We have the probability current in the $x$ direction at each instant
in coordinate time for fixed clock time:

\begin{equation}
j_{\tau}\left({t,x}\right)\equiv-\frac{\imath}{{2m}}\frac{{\partial\psi_{\tau}^{*}\left({t,x}\right)}}{{\partial x}}{\psi_{\tau}}\left({t,x}\right)+\frac{\imath}{{2m}}\psi_{\tau}^{*}\left({t,x}\right)\frac{{\partial{\psi_{\tau}}\left({t,x}\right)}}{{\partial x}}
\end{equation}

\begin{equation}
j_{\tau}\left({t,x}\right)=\left({\frac{{p_{x}}}{{2m}}\psi_{\tau}^{*}\left({t,x}\right)}\right){\psi_{\tau}}\left({t,x}\right)+\psi_{\tau}^{*}\left({t,x}\right)\frac{{p_{x}}}{{2m}}{\psi_{\tau}}\left({t,x}\right)
\end{equation}

What we are after is the full detection rate at a specific coordinate
time:

\begin{equation}
D\left(t\right)=\int{d\tau{D_{\tau}}\left(t\right)}
\end{equation}
 If we accept that the equality in equation \ref{eq:probability-current-4d}
applies at each instant in coordinate time (detailed balance) we have:
\begin{equation}
{D_{\tau}}\left(t\right)={\left.{{j_{\tau}}\left({t,x}\right)}\right|_{x=0}}
\end{equation}
giving:

\begin{equation}
D\left(t\right)={\left.{\int{d\tau{j_{\tau}}\left({t,x}\right)}}\right|_{x=0}}\label{eq:detection-4d}
\end{equation}

Recall that the duration in clock time of a path represents the length
of the path -- in the discrete case the number of steps. To get the
sum of all paths ending at a specific coordinate time, we need to
sum over all possible path lengths. That is this integral.

As in the SQM we are assuming that once the paths enter the detector
they do not return. And as with SQM, this means we have at best a
reasonable first approximation. However as the SQM approximation works
well in practice and as we are interested only in first corrections
to SQM, this should be enough to achieve falsifiability.

\subsubsection{Application to a Gaussian test function}

\label{subsec:tqm-detection-gtf}

If we construct the wave function as the direct product of time and
space parts:

\begin{equation}
{\varphi_{\tau}}\left({t,x}\right)={{\tilde{\varphi}}_{\tau}}\left(t\right){{\bar{\varphi}}_{\tau}}\left(x\right)
\end{equation}

then the expression for the probability current simplifies:

\begin{equation}
{j_{\tau}}\left({t,x}\right)=\left({\left({\frac{{p_{x}}}{{2m}}\bar{\psi}_{\tau}^{*}\left(x\right)}\right){{\bar{\psi}}_{\tau}}\left(x\right)+\bar{\psi}_{\tau}^{*}\left(x\right)\frac{{p_{x}}}{{2m}}{{\bar{\psi}}_{\tau}}\left(x\right)}\right)\tilde{\varphi}_{\tau}^{*}\left(t\right){{\tilde{\varphi}}_{\tau}}\left(t\right)
\end{equation}
or:

\begin{equation}
{D_{\tau}}\left(t\right)={{\bar{D}}_{\tau}}{{\tilde{\rho}}_{\tau}}\left(t\right)
\end{equation}

Using the time wave function from equation \ref{eq:wf-time-tau} we
have for the non-relativistic probability density in time:

\begin{equation}
\tilde{\rho}_{\tau}\left(t\right)=\sqrt{\frac{1}{\pi\sigma_{t}^{2}\left|f_{\tau}^{\left(t\right)}\right|^{2}}}e^{-\frac{1}{\sigma_{t}^{2}\left|f_{\tau}^{\left(t\right)}\right|^{2}}\left(t-\tau\right)^{2}}
\end{equation}

To get the probability for detection at coordinate time $t$ we convolute
clock time with coordinate time. To solve we look at the limit in
long time (as with equation \ref{eq:probdens-space}): 

\begin{equation}
\sigma_{t}^{2}\left|f_{\tau}^{\left(t\right)}\right|^{2}=\sigma_{t}^{2}+\frac{\tau^{2}}{m^{2}\sigma_{t}^{2}}=\sigma_{t}^{2}+\frac{\left(\bar{\tau}+\delta\tau\right)^{2}}{m^{2}\sigma_{t}^{2}}\approx\frac{\left(\bar{\tau}+\delta\tau\right)^{2}}{m^{2}\sigma_{t}^{2}}\approx\frac{\bar{\tau}^{2}}{m^{2}\sigma_{t}^{2}}
\end{equation}

This gives the probability density:

\begin{equation}
{{\tilde{\rho}}_{\tau}}\left(t\right)\approx\sqrt{\frac{1}{{\pi\tilde{\sigma}_{\tau}^{2}}}}{e^{-\frac{1}{{\tilde{\sigma}_{\tau}^{2}}}{{\left({t-\tau}\right)}^{2}}}},{{\tilde{\sigma}}_{\tau}}\equiv\frac{{\bar{\tau}}}{{m{\sigma_{t}}}}
\end{equation}

The convolution over $\tau$ is trivial giving: 
\begin{equation}
{{\rho_{\bar{\tau}}}\left(t\right)=\sqrt{\frac{1}{{\pi\sigma_{\tau}^{2}}}}{e^{-\frac{{{\left({t-\bar{\tau}}\right)}^{2}}}{{\sigma_{\tau}^{2}}}}}}
\end{equation}
with the total dispersion in clock time being the sum of the dispersions
in coordinate time and in space: 
\begin{equation}
{\sigma_{\tau}^{2}\equiv\tilde{\sigma}_{\tau}^{2}+\bar{\sigma}_{\tau}^{2}}
\end{equation}
This is reasonably simple. The uncertainty is: 
\begin{equation}
\Delta\tau=\frac{1}{{\sqrt{2}}}\sqrt{\tilde{\sigma}_{\tau}^{2}+\bar{\sigma}_{\tau}^{2}}\label{eq:total-uncertainty-in-time}
\end{equation}
We collect the definitions for the two dispersions:
\begin{equation}
\begin{array}{c}
\bar{\sigma}_{\tau}^{2}=\frac{{{\bar{\tau}}^{2}}}{{{m^{2}}{v^{2}}\sigma_{x}^{2}}}\hfill\\
\tilde{\sigma_{t}}^{2}\approx\frac{{{\bar{\tau}}^{2}}}{{{m^{2}}\sigma_{t}^{2}}}\hfill
\end{array}\label{eq:tqm-dispersions}
\end{equation}

As noted, we expect the particle wave functions to have initial dispersions
in energy/time comparable to their dispersions in momentum/space:

\begin{equation}
\sigma_{t}\sim\sigma_{x}
\end{equation}

In the non-relativistic case, $v\ll1$ the total uncertainty will
be dominated by the space part. While we have looked specifically
at wave functions composed of a direct product of GTFs in time and
space, the underlying arguments are general. Therefore we expect that
in the case of non-relativistic particles, the uncertainty in time-of-arrival
will show evidence of dispersion in coordinate time, but will be dominated
by the dispersion in space, because this enters with a $\frac{1}{v}$
factor and in the non-relativistic case $v$ is by definition small.

\section{Applications}

\paragraph{Single slit in time}

At non-relativistic velocities, we expect that the uncertainty in
time-of-arrival will be dominated by the space part. To increase the
effect of uncertainty in time we can run the wave function through
a single slit in time, i.e. a camera shutter which opens and closes
very rapidly. In SQM, the wave function will merely be clipped and
the uncertainty in time at the detector correspondingly reduced. But
in TQM the reduction in $\Delta t$ will increase $\Delta E$. The
increase in uncertainty in energy will cause the wave function to
be more spread out in time, making the uncertainty in time at the
detector arbitrarily great. The case is analogous to the corresponding
single slit in space, with $\Delta E,\Delta t$ substituting for $\Delta p,\Delta x$.

In our previous paper we examined this case, but our analysis took
as its starting point the Kijowski metric and associated approaches.
In particular, we shifted back and forth between $q$ and $p$ in
phase space in a way that we now recognize as suspect. 

This obscured an important subtlety in the case of non-relativistic
particles. Consider a non-relativistic particle going through a time
gate. If the gate is open for only a short time, the particle must
have come from an area close to the time gate. Therefore its position
in space is also fairly certain. If the particle is traveling with
non-relativistic velocity $v\ll1$ then $\Delta x\sim v\Delta t\Rightarrow\Delta x\ll\Delta t$.
This in turn drives up the uncertainty in momentum, $\Delta p\sim{1\mathord{\left/{\vphantom{1{\Delta x}}}\right.\kern-\nulldelimiterspace}{\Delta x}}$,
leading to increased uncertainty in time at the detector, even for
SQM.

To avoid this problem, we replace the gate with a time-dependent source.
We let the particles propagate for a short time past the source, then
compare the expected uncertainties without and with dispersion in
time. The basic conclusion of the previous work is unchanged: we can
make the difference between the two cases arbitrarily large. We work
out the details in appendix \ref{sec:Single-slit-experiment}.

\paragraph{Double slit in time}

The double slit experiment is often cited as \emph{the} key experiment
for quantum mechanics (mostly famously by Feynman \cite{Feynman:1965ah}).
And Lindner's ``Attosecond Double-Slit Experiment'' \cite{Lindner:2005vv}
provided a key inspiration for the current work, in that it showed
we could now explore interference effects at the necessary attosecond
scale. Building on this, Palacios et al \cite{Palacios:2009vh} have
proposed an experiment using spin-correlated pairs of electrons which
takes advantage of the electrons being indistinguishable to look for
interference in the time-energy domain\footnote{Unfortunately as far as we know this experiment has not yet been performed.}.

Both experiments use the standard quantum theory to do the analysis
of interference in the time/energy dimension. Horwitz \cite{Horwitz:2005ix,Horwitz:2007wb,Horwitz:2015jk}
notes that as time is a parameter in standard quantum mechanics it
is difficult to fully justify this analysis. He provides an alternative
analysis in terms of the Stueckelberg equation (our eq \ref{eq:schroedinger-equation-4d})
which is not subject to this objection. He gets the same spacing for
the fringes as Lindner found.

Unfortunately for this investigation since the fringe spacing is the
same using the Lindner's analysis or the one here the fringe pattern
does not let us distinguish between the two approaches; it contributes
nothing to falsifiability.

We do expect, from the analysis of the single slit experiment, that
the individual fringes will be smoothed out from the additional dispersion
in time: the brights less bright, the darks less dark. However the
contribution of this effect to falsifiability is already covered by
the analysis of the single slit experiment.

\paragraph{Quantum electrodynamics}

While the above is sufficient to establish that we can detect the
effects of uncertainty in time, it is clear that the strongest and
most interesting effects will be seen at high energies and relativistic
speeds, where the effects of quantum field theory (QFT) will need
to be taken into account. 

In the previous paper, we looked at the QFT approach to systems of
spinless, massive particles. We were able to extend TQM to QFT in
a straightforward and unambiguous way. We also showed that we recover
the usual Feynman rules in the appropriate limit. And that we will
see various interesting effects from the additional uncertainty in
time.

One obvious concern was that the additional dimension in the loop
integrals might make the theory still more divergent than QFT already
is, perhaps even unrenormalizable. But in the spin 0 massive case
we do not see the usual ultraviolet (UV) divergences: the combination
of uncertainty in time, entanglement in time, and the use of finite
initial wave functions (rather than plane waves or $\delta$ functions)
suppresses the UV divergences without the need for regularization.

The extension to spin 1/2 particles and to photons appears to be straightforward.
We expect to see the various effects of time-as-observable for spin
1/2 particles and photons: interference in time, entanglement in time,
correlations in time, and so on. We will also look at the implications
for spin correlations, symmetry/anti-symmetry under exchange of identical
particles, and the like.

One caveat: if we analyze the photon propagator using the familiar
trick of letting the photon mass $\mu\to0$\cite{Zee:2010oy} we can
see from inspection of the $\frac{1}{2\mu}$ factor in the propagator
in momentum space:

\begin{equation}
{K_{\tau}}\left({k;k'}\right)=\exp\left({\imath\frac{{{k^{2}}-{\mu^{2}}}}{{2\mu}}\tau}\right){\delta^{4}}\left({k-k'}\right)
\end{equation}

that excursions off-shell will be severely penalized. This in turn
will limit the size of the effects of TQM in experiments such as the
single slit in time. Such experiments are therefore better run with
massive particles, the more massive the better in fact.

In this paper we have estimated the initial wave functions in time
using dimensional and maximum entropy arguments. Once TQM has been
extended to include QED, we should be able to combine the standard
QED results with these to serve as the zeroth term in a perturbation
expansion, then compute first and higher corrections using the standard
methods of time-dependent perturbation theory.

This should open up a wide variety of interesting experiments.

\paragraph{General Relativity}

As noted in the introduction, TQM is a part of the relativistic dynamics
program so can draw on the extensive relativistic dynamics literature.
In particular we can take advantage of Horwitz's extension of relativistic
dynamics to General Relativity for single particles, the classical
many body problem, and the quantum many body problem \cite{Horwitz:2018aa,Horwitz:2020vh,Horwitz:2021wd}.

There would appear to be an interesting complementarity between the
work of Horwitz and this work. First, TQM supplies an explicit source
for the invariant monotonic parameter $\tau$ so that we do not need
to introduce this as an additional assumption. Second, TQM avoids
the UV divergences which have created significant difficulties for
work to extend QFT to General Relativity. Therefore an appropriate
combination of the two approaches might provide useful insights into
the extension of QFT to General Relativity.

\section{Discussion}

\label{sec:discussion}
\begin{quotation}
“It is not surprising that our language should be incapable of describing
the processes occurring within the atoms, for, as has been remarked,
it was invented to describe the experience of daily life, and these
consist only of processes involving exceedingly large numbers of atoms.
Furthermore, it is very difficult to modify our language so that it
will be able to describe these atomic processes, for words can only
describe things of which we can form mental pictures, and this ability,
too, is a result of daily experience. Fortunately, mathematics is
not subject to this limitation, and it has been possible to invent
a mathematical scheme --- the quantum theory --- which seems entirely
adequate for the treatment of atomic processes…” -- Werner Heisenberg
\cite{Heisenberg:1930kb}
\end{quotation}

\subparagraph{Falsifiability}

The Heisenberg uncertainty principle (HUP) in time follows directly
from quantum mechanics and relativity. This was clear to both Einstein
and Bohr in 1930. However quantum mechanics since then has not in
general included it. Given that the relevant attosecond time scales
have been too small to work with -- until recently -- this is reasonable
enough.

However it is now possible to look at the time at the relevant scales;
it is therefore necessary to do so, if we are understand either time
or quantum mechanics. The specific predictions made here are based
on only covariance and the requirement of consistency with existing
results, they are by construction without free parameters. They should
therefore provide at a minimum a reasonable order-of-magnitude target.

The time-of-arrival experiment we have focused on here provides merely
the most obvious line of attack. Essentially any experiment looking
at time-dependent phenomena is a potential candidate. For instance
in Auletta \cite{Auletta:2000vj} there are about three hundred foundational
experiments; most can be converted into a test of uncertainty in time
by replacing an $x$ with a $t.$ Examples include the single slit
in time, the double slit in time, and so on. One may also look at
variations intended to multiply normally small effects, as experiments
using resonance or diffraction effects.

\subparagraph{Implications of negative result striking}

From symmetry, one may argue that a positive result is the more likely,
but even a negative result would be interesting.

For instance, assume we have established there is no HUP in time in
one frame. Consider the same experiment from another frame moving
at high velocity relative to the first. Consider say a Gaussian test
function (GTF) in $x$ in the first; make a Lorentz boost of this
into the second frame. The Lorentz boost will transform $x\to\gamma\left(x-v\tau\right)$.
This will turn the uncertainty in space into a mixture of uncertainty
in space and in time. We then look for uncertainty in time in the
boosted frame.

If the HUP in time is also rejected in the second, how do we maintain
the principle of covariance? If the HUP in time is present in the
second frame, then we can defined a preferred frame as the one in
which HUP in time is maximally falsified. Such a preferred frame is
anathema to general relativity.

Therefore exploring the precise character of the negative result --
uniform across frames, more in some frames than others, and so on
-- would itself represent an interesting research program.

\subparagraph{Positive result would have significant practical applications}

If on the other hand, the wave function extends in time, this would
not only be interesting from the point of view of fundamental physics,
it would open up a variety of practical applications. For instance,
there would be an additional channel for quantum communication systems
to use. Memristors and other time-dependent circuit elements would
show interesting effects. In attosecond chemistry and biochemistry
we would expect to see forces of anticipation and regret; if the wave
function extends in time, it will cause interactions to start earlier
and end later than would otherwise be the case. The mysteries of protein
folding could be attacked from a fresh perspective and perhaps unexpected
temporal subtleties found.

The applications in quantum computing are particularly interesting.
Quantum computers will need to compensate for the effects of decoherence
along the time dimension. But they should also be able to take advantage
of additional computing opportunities along the time dimension. And
if we have a deeper understanding of the relationship between time
and measurement, we may find opportunities to ``cut across the diagonal''
in the design of quantum computers.

\ack

I thank my long time friend Jonathan Smith for invaluable encouragement,
guidance, and practical assistance.

I thank Ferne Cohen Welch for extraordinary moral and practical support.

I thank Martin Land, L. P. Horwitz, and the other organizers and participants
of the International Association for Relativistic Dynamics (IARD)
2020 Conference for encouragement, useful discussions, and hosting
a talk on this paper at the IARD 2020 conference.

I thank the reviewer who drew my attention to Horwitz's \cite{Horwitz:2018aa}.

I thank Steven Libby for several useful conversations. 

I thank Larry Sorensen for many helpful references. I thank Ashley
Fidler for helpful references to the attosecond physics literature.

And I thank Avi Marchewka for an interesting and instructive conversation
about various approaches to time-of-arrival measurements.

And I note none of the above are in any way responsible for any errors
of commission or omission in this work.

.

\appendix

\section{Direct computation of the detection rate in diffusion }

\label{sec:diffusion-detection}

We use conservation of probability and the method of images to make
a direct computation of the detection rate in the case of diffusion.
The approach is similar to the one used by Marchewka and Schuss to
compute the probability current (subsection \ref{subsec:Probability-current-sqm}),
although the context is classical rather than quantum.

Define ${G_{\tau}}\left(x\right)$ as the probability to get from
$d\to x$ without touching the detector at $0$. We get $G$ from
the method of images by the same logic as in the discrete case:

\begin{equation}
{G_{\tau}}\left(x\right)={P_{\tau}}\left({x;-d}\right)-{P_{\tau}}\left({x;d}\right)
\end{equation}

$G$ obeys the diffusion equation:

\begin{equation}
\frac{{\partial{G_{\tau}}}}{{\partial\tau}}=\frac{1}{{2m}}\frac{{\partial^{2}}}{{\partial{x^{2}}}}{G_{\tau}}
\end{equation}

Take $D_{\tau}$ as the rate of detection at $\tau$. From conservation
of probability we have:

\begin{equation}
1=\int\limits _{-\infty}^{0}{dx{G_{\tau}}\left(x\right)}+\int\limits _{0}^{\tau}{d\tau'{D_{\tau'}}}
\end{equation}

Take the derivative with respect to $\tau$ and apply the diffusion
equation:

\begin{equation}
{D_{\tau}}=-\int\limits _{-\infty}^{0}{dx\frac{{\partial{G_{\tau}}}}{{\partial\tau}}\left(x\right)}=-\int\limits _{-\infty}^{0}{dx\frac{1}{{2m}}\frac{{\partial^{2}}}{{\partial{x^{2}}}}G}
\end{equation}

Since the term on the right is the integral of a derivative we have:

\begin{equation}
{D_{\tau}}=-\frac{1}{{2m}}{\left.{\frac{\partial}{{\partial x}}G}\right|_{x=0}}
\end{equation}

By using the explicit form for $P$ (equation \ref{eq:diffusion-probability})
we get:

\begin{equation}
{D_{\tau}}=\frac{d}{\tau}{P_{\tau}}\left(d\right)
\end{equation}

as above (equation \ref{eq:diffusion-detection}).

\section{Alternate derivation of the detection amplitude}

\label{sec:Alternate-derivation}

\begin{figure}
\includegraphics[scale=0.67]{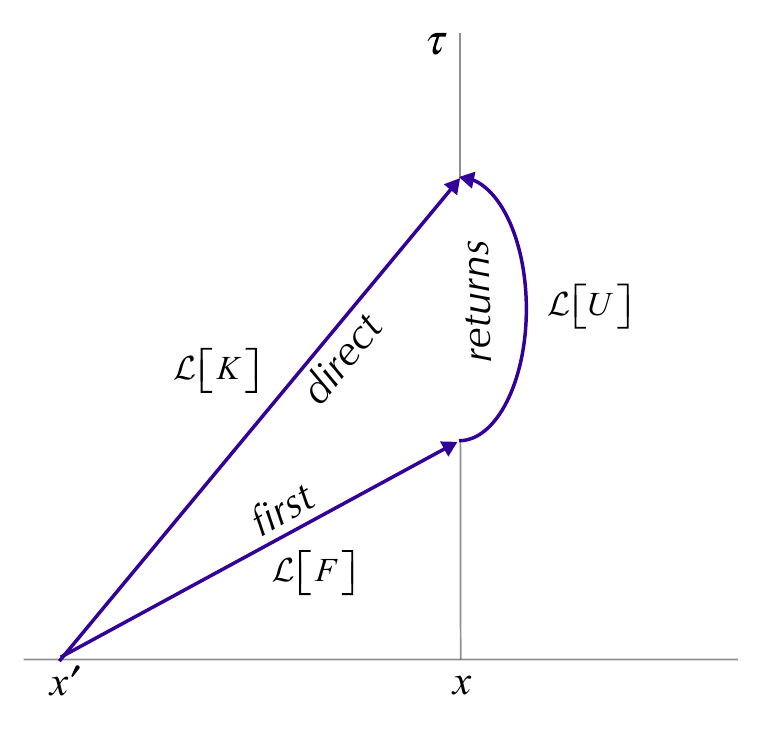}

\caption{Computation of amplitude for first arrival by using the Laplace transform}
\end{figure}

We compute $F$ by using the Laplace transform.

To get to any specific point $x$ we have to get to the point the
first time, then return to it zero or more times. Therefore we can
write the full kernel $K$ as the convolution of the kernel $F$ to
arrive for the first time and the kernel $U$ to return zero or more
times:

\begin{equation}
{K_{\tau}}\left({x;x'}\right)=\int\limits _{0}^{\tau}{d\tau'{U_{\tau'}}\left(x\right){F_{\tau'}}\left({x;x'}\right)}
\end{equation}

Since the kernel is invariant under space translation:

\begin{equation}
\begin{gathered}{K_{\tau}}\left({x;x'}\right)={K_{\tau}}\left({x-x'}\right)\hfill\\
{F_{\tau'}}\left({x;x'}\right)={F_{\tau}}\left({x-x'}\right)\hfill
\end{gathered}
\end{equation}

and:

\begin{equation}
{U_{\tau}}\left(x\right)={U_{\tau}}\left(0\right)={K_{\tau}}\left(0\right)
\end{equation}

we can simplify the convolution to:

\begin{equation}
{K_{\tau}}\left({x-x'}\right)=\int\limits _{0}^{\tau}{d\tau'{U_{\tau'}}F_{\tau'}\left({x-x'}\right)}
\end{equation}

The Laplace transform of $K$ is the product of the Laplace transform
of $U$ and $F$:

\begin{equation}
\mathcal{L}\left[K\right]=\mathcal{L}\left[U\right]\mathcal{L}\left[F\right]
\end{equation}

The free kernel is:

\begin{equation}
{K_{\tau}}\left(x\right)=\sqrt{\frac{m}{{2\pi\imath\tau}}}\exp\left({\imath m\frac{{x^{2}}}{{2\tau}}}\right)\label{eq:free-kernel-in-space}
\end{equation}

with Laplace transform:

\begin{equation}
\mathcal{L}\left[K\right]=-\left({-{1^{3/4}}}\right)\frac{{e^{\left({-1+\imath}\right)\sqrt{ms}\left|x\right|}}}{{\sqrt{2}\sqrt{s}}}\sqrt{m}
\end{equation}

To get the Laplace transform of $U$ take the $x=0$ case:

\begin{equation}
\mathcal{L}\left[U\right]=-\frac{{\left({-{1^{3/4}}}\right)}}{{\sqrt{2}\sqrt{s}}}\sqrt{m}
\end{equation}
 Therefore we have the Laplace transform for $F$:

\begin{equation}
\mathcal{L}\left[F\right]={e^{\left({-1+\imath}\right)\sqrt{ms}\left|x\right|}}
\end{equation}

and we get $F_{\tau}$ from the inverse Laplace transform:

\begin{equation}
\begin{gathered}\hfill\\
{F_{\tau}}=\frac{\left|x\right|}{\tau}{K_{\tau}}\left(x\right)\hfill
\end{gathered}
\end{equation}

\section{Single slit in time}

\label{sec:Single-slit-experiment}

We compare the results of the single slit in time in SQM and TQM.
We model the single slit in time as a particle source located at $x=-d$,
emitting particles with momentum $p_{\ensuremath{0}}$ in the $x$-direction,
velocity ${v_{0}}=\frac{{p_{0}}}{m}$. 

In the SQM case, we assume that the wave function is emitted with
probability $\bar{G}_{\tau}$:

\begin{equation}
\bar{G}\left(\tau\right)\equiv\frac{1}{{\sqrt{2\pi{W^{2}}}}}\exp\left({-\frac{{\tau^{2}}}{{2{W^{2}}}}}\right)
\end{equation}

This is normalized to one, with uncertainty in time $\Delta\tau=W$. 

To extend to TQM we will replace this probability with an amplitude:

\begin{equation}
{{\tilde{\varphi}}_{0}}\left(t_{0}\right)=\sqrt[4]{{\frac{1}{{\pi\sigma_{t}^{2}}}}}\exp\left({-\imath E_{0}t_{0}-\frac{{t_{0}^{2}}}{{2\sigma_{t}^{2}}}}\right)\label{eq:single-wf-time-initial}
\end{equation}

This has probability distribution and uncertainty:

\begin{equation}
\begin{gathered}{{\tilde{\rho}}_{0}}\left({t_{0}}\right)=\sqrt{\frac{1}{{\pi\sigma_{t}^{2}}}}\exp\left({-\frac{{t_{0}^{2}}}{{\sigma_{t}^{2}}}}\right)\\
\Delta t=\frac{{\sigma_{t}}}{{\sqrt{2}}}
\end{gathered}
\end{equation}

We will take ${\sigma_{t}}=\sqrt{2}W$ so that the uncertainty from
the gate is equal to the uncertainty from the initial distribution
in coordinate time, to make the comparison as fair as possible.

The detector will be positioned at $x=0$. The average time of arrival
is $\bar{t}=\bar{\tau}=\frac{d}{{v_{0}}}$. We are interested in the
uncertainty in time at the detector as defined above for SQM and TQM. 

\subsection{Single slit in time in SQM}

\begin{figure}
\includegraphics[scale=0.75]{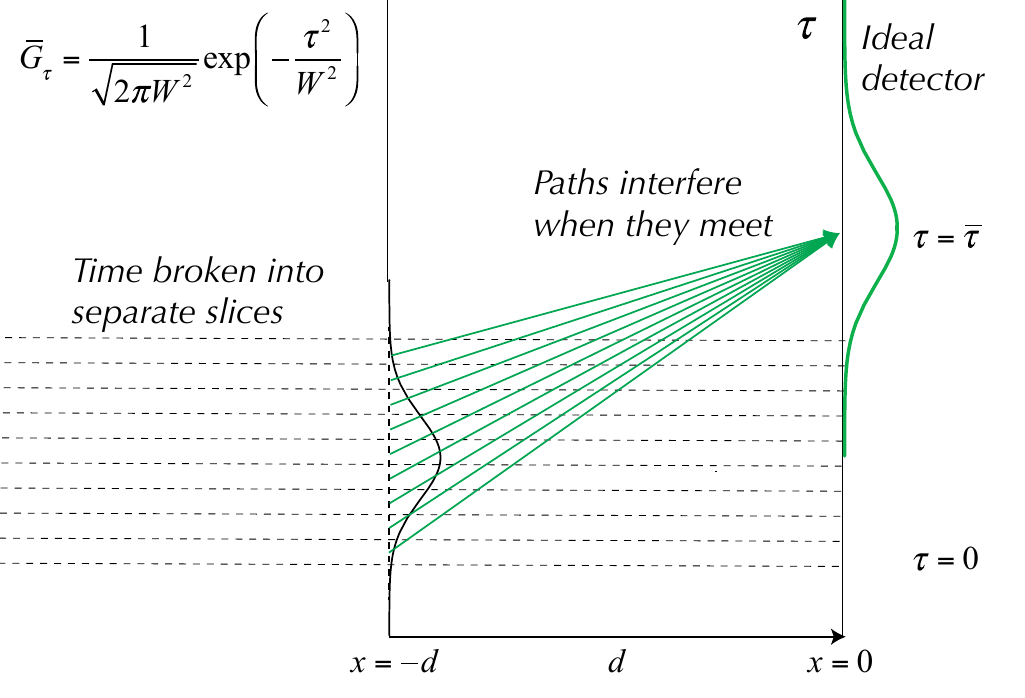}

\caption{Single slit in time in SQM}
\end{figure}

In SQM the wave function extends in only the three space dimensions.
If we break time up into slices, there is no interference across slices.
At each time slice, there is an amplitude for paths from that slice
to get to the detector. But if we look at the individual paths emitted
during that time slice, some get to the detector at one time, some
at an earlier or a later time. At the detector all paths that arrive
at the same clock time interfere, constructively or destructively,
as may happen. This is the picture used by Lindner et al in their
analysis. The peaks in the incoming electric wavelet ejected electrons
at different times, but when the electrons ejected from different
times arrived at the detector at the same time they interfered.

To make this specific, we will use a simple model.

For convenience, we will assume that the source has an overall time
dependency of $\exp\left({-\imath\frac{{p_{0}^{2}}}{{2m}}{\tau_{G}}}\right)$
so that our initial wave function is:

\begin{equation}
{{\bar{\varphi}}_{G}}\left({x_{G}}\right)=\sqrt[4]{{\frac{1}{{\pi\sigma_{x}^{2}}}}}{e^{\imath{p_{0}}{x_{G}}-\frac{1}{{2\sigma_{x}^{2}}}x_{G}^{2}-\frac{{p_{0}^{2}}}{{2m}}{\tau_{G}}}}
\end{equation}

This is normalized to one at the start:

\begin{equation}
1=\int{d{x_{G}}\varphi_{G}^{*}\left({x_{G}}\right){\varphi_{G}}\left({x_{G}}\right)}
\end{equation}

So the particle will have a total probability of being emitted of
one:

\begin{equation}
1=\int\limits _{-\infty}^{\infty}{d\tau{{\bar{G}}_{\tau}}\int{d{x_{G}}\varphi_{G}^{*}\left({x_{G}}\right){\varphi_{G}}\left({x_{G}}\right)}}
\end{equation}

The amplitude at the detector from a single moment at the gate will
be given by:

\begin{equation}
{{\bar{\varphi}}_{DG}}\left(x\right)=\sqrt[4]{{\frac{1}{{\pi\sigma_{x}^{2}}}}}\sqrt{\frac{1}{{f_{DG}^{\left(x\right)}}}}{e^{\imath{p_{0}}x-\frac{1}{{2\sigma_{x}^{2}f_{DG}^{\left(x\right)}}}{{\left({x+d-{v_{0}}{\tau_{DG}}}\right)}^{2}}-\imath\frac{{p_{0}^{2}}}{{2m}}{\tau_{D}}}}
\end{equation}

with ancillary definitions:

\begin{equation}
\begin{gathered}{\tau_{DG}}\equiv{\tau_{D}}-{\tau_{G}}\hfill\\
f_{DG}^{\left(x\right)}=1+\imath\frac{{\tau_{DG}}}{{m\sigma_{x}^{2}}}\hfill
\end{gathered}
\end{equation}

We have:

\begin{equation}
x+d-{v_{0}}\left({{\tau_{D}}-{\tau_{G}}}\right)=-{v_{0}}\left({\delta\tau-{\tau_{G}}}\right)
\end{equation}

Both $\delta\tau$ and $\tau_{G}$ are expected small. We can therefore
justify taking:

\begin{equation}
f_{DG}^{\left(x\right)}=1+\imath\frac{{\tau_{DG}}}{{m\sigma_{x}^{2}}}\approx f_{\bar{\tau}}^{\left(x\right)}=1+\imath\frac{{\bar{\tau}}}{{m\sigma_{x}^{2}}}
\end{equation}

Giving:

\begin{equation}
{{\bar{\varphi}}_{DG}}\left(x\right)=\sqrt[4]{{\frac{1}{{\pi\sigma_{x}^{2}}}}}\sqrt{\frac{1}{{f_{\vec{\tau}}^{\left(x\right)}}}}{e^{\imath{p_{0}}x-\frac{{v_{0}^{2}}}{{2\sigma_{x}^{2}f_{\vec{\tau}}^{\left(x\right)}}}{{\left({\delta\tau-{\tau_{G}}}\right)}^{2}}-\imath\frac{{p_{0}^{2}}}{{2m}}{\tau_{D}}}}
\end{equation}

To get the full wave function at the detector we need to take the
convolution of this with the gate function:

\begin{equation}
{{\bar{\psi}}_{D}}\left(x\right)=\int\limits _{-\infty}^{\infty}{d{\tau_{G}}\bar{G}\left({\tau_{G}}\right){{\bar{\varphi}}_{DG}}\left(x\right)}
\end{equation}

giving:

\begin{equation}
{{\bar{\psi}}_{D}}\left(x\right)=\frac{{\sigma_{x}}}{{v_{0}}}\sqrt[4]{{\frac{1}{{\pi\sigma_{x}^{2}}}}}\frac{1}{{\sqrt{2\pi\left({\frac{{\sigma_{x}^{2}}}{{v_{0}^{2}}}+{W^{2}}+\imath\frac{{\bar{\tau}}}{{mv_{0}^{2}}}}\right)}}}\exp\left({\imath{p_{0}}x-\frac{{{\left({\delta\tau}\right)}^{2}}}{{2\left({\frac{{\sigma_{x}^{2}}}{{v_{0}^{2}}}+{W^{2}}+\imath\frac{{\bar{\tau}}}{{mv_{0}^{2}}}}\right)}}-\imath\frac{{p_{0}^{2}}}{{2m}}{\tau_{D}}}\right)
\end{equation}

We can see that the effect of the gate is to increase the effective
dispersion in space:

\begin{equation}
\frac{{\sigma_{x}^{2}}}{{v_{0}^{2}}}\to\frac{{\sigma_{x}^{2}}}{{v_{0}^{2}}}+{W^{2}}\Rightarrow\sigma_{x}^{2}\to\Sigma_{x}^{2}\equiv\sigma_{x}^{2}+v_{0}^{2}{W^{2}}
\end{equation}

The gate effectively adds an uncertainty of $vW$ to the original
uncertainty in space. As $vW$ is about the distance a particle would
cross while the gate is open, this is reasonable.

We get then as the associated uncertainty in time (eq: \ref{eq:sqm-uncertainty}):

\[
\Delta\tau=\frac{1}{{\sqrt{2}}}{{\bar{\Sigma}}_{\tau}}=\frac{1}{{\sqrt{2}}}\frac{1}{{m{v_{0}}{\Sigma_{x}}}}\bar{\tau}
\]

The longer the gate stays open the greater the resulting uncertainty
in time at the detector. The shorter the gate stays open the less
the uncertainty in time at the detector, with the minimum being the
uncertainty for a free GTF released at the time and location of the
gate.

\subsection{Single slit in time in TQM}

\begin{figure}
\includegraphics[scale=0.75]{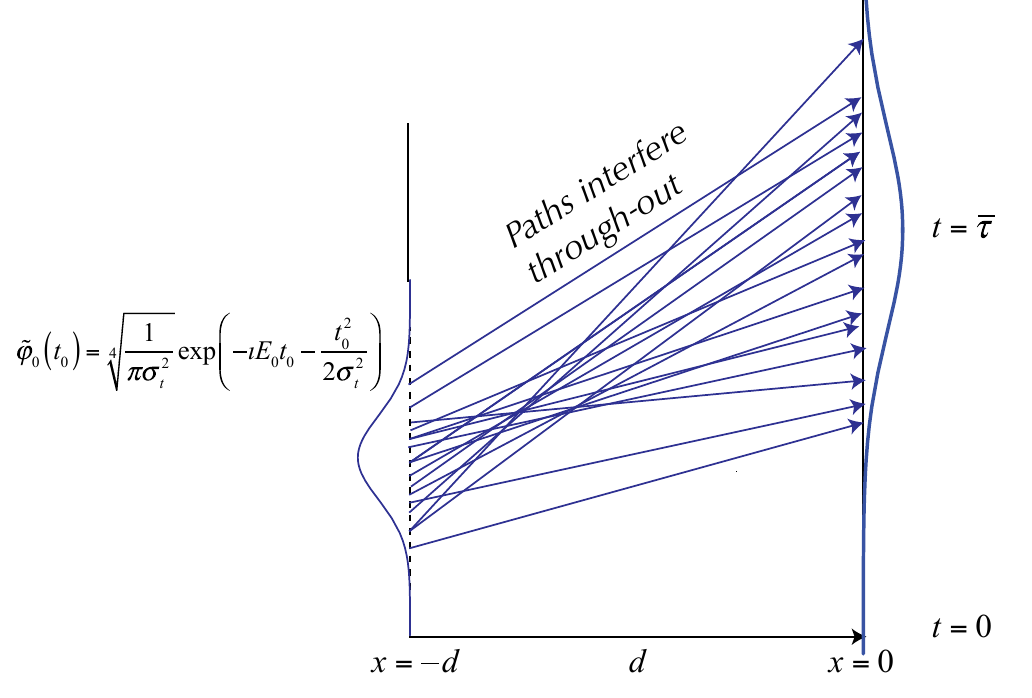}

\caption{Single slit in time in TQM}
\end{figure}

In TQM the wave function extends in all four dimensions. There is
interference across time so it is no longer legitimate to break time
up into separate slices (except for purposes of analysis of course).
We will take the source as centered at a specific moment in coordinate
time, $t=0$. 

We can write the four dimensional wave function as the product of
the time (eq: \ref{eq:single-wf-time-initial}) and space parts:

\begin{equation}
{\varphi_{0}}\left({{t_{0}},{{\vec{x}}_{0}}}\right)={\tilde{\varphi}_{0}}\left({t_{0}}\right){\bar{\varphi}_{0}}\left({{\vec{x}}_{0}}\right)
\end{equation}

The space part is as above for clock time $\tau=0$.

The time part at clock time $\tau$ is:
\begin{quotation}
\begin{equation}
{{\tilde{\varphi}}_{\tau}}\left(t\right)=\sqrt[4]{{\frac{1}{{\pi\sigma_{t}^{2}}}}}\sqrt{\frac{1}{{f_{\tau}^{\left(t\right)}}}}{e^{-\imath{E_{0}}t-\frac{1}{{2\sigma_{t}^{2}f_{\tau}^{\left(t\right)}}}{{\left({t-\frac{{E_{0}}}{m}\tau}\right)}^{2}}}}
\end{equation}
\end{quotation}
For a non-relativistic particle $\frac{E}{m}\approx1$. The treatment
in subsection \ref{subsec:tqm-detection-gtf} applies giving the uncertainty
in time at the detector as:

\begin{equation}
\Delta\tau=\frac{1}{{\sqrt{2}}}\sqrt{\tilde{\sigma}_{\tau}^{2}+\bar{\sigma}_{\tau}^{2}}
\end{equation}

with contributions from the space and time parts of:

\begin{equation}
\begin{gathered}\bar{\sigma}_{\tau}^{2}=\frac{{{\bar{\tau}}^{2}}}{{{m^{2}}{v^{2}}\sigma_{x}^{2}}}\hfill\\
\tilde{\sigma}_{\tau}^{2}=\frac{{{\bar{\tau}}^{2}}}{{{m^{2}}\sigma_{t}^{2}}}=\frac{{{\bar{\tau}}^{2}}}{{2{m^{2}}{W^{2}}}}\hfill
\end{gathered}
\end{equation}

So we have the scaled uncertainty in time as:

\begin{equation}
\frac{{\Delta\tau}}{{\bar{\tau}}}=\frac{1}{{\sqrt{2}}}\frac{1}{m}\sqrt{\frac{1}{{{v^{2}}\sigma_{x}^{2}}}+\frac{1}{{2{W^{2}}}}}
\end{equation}

We can see that when $W\sim v\sigma_{x}$ the effects of dispersion
in time will be comparable to those from dispersion in space. And
as $W\to0$, the uncertainty in time at the detector will be dominated
by the width of the gate in time, going as $\frac{1}{W}$.

So the intrinsic uncertainty in space of a GTF creates a corresponding
minimum uncertainty in time at the detector given by $\bar{\sigma}_{\tau}$.
In SQM the effects of the gate drop to zero as $W\to0$ while in TQM
they go to infinity. In SQM the wave function is clipped in time;
in TQM it is diffracted.

We therefore have the unambiguous signal -- even in the non-relativistic
case -- that we need to achieve practical falsifiability.

\bibliography{toa}
\bibliographystyle{iopart-num}

\end{document}